\documentclass[twocolumn,prb, showpacs,showkeys]{revtex4-1}
\usepackage{hyperref}
\usepackage[dvipdfm]{graphicx}
\usepackage{amsmath}
\usepackage{amssymb}
\usepackage{amsfonts}
\usepackage{booktabs}    
\usepackage{subfigure}
\usepackage{longtable}
\usepackage{enumerate}
\usepackage{multirow}
\usepackage{color}
\usepackage{braket}

\begin{document}
\title{One-dimensional physics in transition-metal nanowires:\\
Renormalization group and bosonization analysis}
\author{Jun-ichi Okamoto}
\email{okamoto@phys.columbia.edu}
\affiliation{Department of Physics, Columbia University, 538 West 120th Street, New York, New York 10027, USA}

\author{A. J. Millis}
\email{millis@phys.columbia.edu}
\affiliation{Department of Physics, Columbia University, 538 West 120th Street, New York, New York 10027, USA}

\date{\today}

\begin{abstract}
We study the one-dimensional two-orbital Hubbard model with general local interactions including a pair-hopping term. The model might be realized in one-dimensional transition-metal nanowires. Phase diagrams at $T=0$ are obtained by numerical integration of renormalization group equations and bosonization. Particular attention is paid to the effects of orbital degeneracy (or near-degeneracy), interactions favoring locally high-spin configurations, and velocity differences. Dynamical symmetry enlargement and duality approaches are employed to determine ground states and to understand quantum phase transitions between them. An important result is that the pair-hopping term and associated orbital symmetry can lead to new insulating states. The ground state for spin-polarized case is also discussed.
\end{abstract}


\maketitle

\section{Introduction}
\label{intro}
Theoretical models involving two orbitals per unit cell in one dimension have been employed to understand various strongly correlated systems e.g. heavy fermion compounds,\cite{Varma1985a, Strong1994a, Fujimoto1994} high-$T_c$ superconductors,\cite{Fabrizio1992,Finkelstein1993,Khveshchenko1994d,Balents1996,Schulz1996,Shelton1996a,Lin1998,Lee2005,Chudzinski2008} spin-ladders,\cite{Shelton1996b, Kim2000} and Hubbard-ladders.\cite{Azaria1999,Tsuchiizu2002,Wu2003} These models are often referred to as ``ladder models'' because one may visualize the two states as opposite sides of a rung. Recent experiments demonstrate the formation of self-assembled transition-metal nanowires, adding another possible realization of ladder models.\cite{Wang2008, Zaki2009} In such systems, one-dimensional (1D) nanowires composed of transition-metal atoms are confined at step edges on substrate surface. In many physically relevant cases, the surface bandgap structure of the substrate material is such that the electronic states of the adatoms are decoupled from the bulk substrate bands (at least to leading order and for low energies) and the  $d$-orbital bands derived from the transition-metal nanowire form a multi-component one-dimensional Fermi gas with variety of interactions. Classifying the kinds of behavior which may be observed in these systems is an important open question. In this paper, we will focus on the case where only two bands cross the Fermi energy. This situation is equivalent to two-leg ladder models, which have been previously studied.

Past studies revealed that two-leg ladder systems may exhibit dynamical symmetry enlargement (DSE) \cite{Lin1998,Konik2002,Controzzi2005}and dualities among ground states.\cite{Boulat2009,Nonne2010} DSE occurs when a RG flow leads to an effective low energy fixed point which exhibits a higher symmetry than that of original lattice Hamiltonian. In two-leg ladder models, the emergent symmetry is known to be O(6)$\times$U(1) when the system is away from half filling, and O(8) at half filling. The different ground states of these low-energy theories are related to each other by duality mappings, which are generalizations of the Kramers-Wannier duality seen in the two dimensional Ising model.\cite{Kramers1941}

We expect that many qualitative aspects of two-leg ladder models, including DSE and duality properties, hold also in our models of transition-metal nanowires. However, new features of transition-metal wires require additional investigation. The orbital degeneracy of the transition-metal $d$-levels permits a rich set of onsite Coulomb interactions. In particular, the Hund coupling favors locally high spin configurations, and we speculate that the ground states may exhibit non-zero spin structure. Furthermore, the pair-hopping term, which has been neglected in some previous studies, is found to be important; it implies the system has a unique orbital symmetry. We generically expect that orbital symmetry breaking at the level of the one-body terms leads to velocity differences between different orbitals. A large velocity difference complicates the application of established methods such as refermionization\cite{Tsvelik2011} to our model. Strong correlation may also lead to spin-polarized states or to ferromagnetism observed, for example, in  Co nanowires.\cite{Gambardella2002a} Finally, the possible realization of an orbital selective Mott phase\cite{Koga2004} in one-dimension is an open issue.

In order to understand the effect of these features of transition-metal nanowires, we study the one-dimensional two-orbital Hubbard model using perturbative renormalization group (RG) and bosonization approaches. We combine the ideas of DSE and duality to list the possible ground states of our models. We find the form of interaction relevant to transition-metal $d$-levels leads to a new group of 8 insulating phases when the two orbitals are completely degenerate. We then obtain ground state phase diagrams using RG and bosonization. In physically relevant parameter regimes, the stability of the ground states to velocity differences is also investigated. To be complete, the fully spin-polarized case, where the model is reduced to a Hubbard model with or without magnetic field, is also briefly analyzed.

The methods we employ in this paper are strictly applicable when the system is truly one-dimensional, and the interaction is weak. Our results are complimentary to those we previously obtained for the same model using mean-field approaches.\cite{Okamoto2011} The present results allow more complete understanding of the weak-coupling regime, but the mean-field theory can treat the intermediate and strong coupling regimes.

The organization of this paper is as follows. In Sec. \ref{model} we explain the model and approximation we employed. In Sec. \ref{bosonization}, we derive bosonized forms of the model. Section \ref{order parameters} is devoted to explanation of various possible order parameters. In section \ref{duality}, we use the idea of dynamical symmetry enlargement and duality to understand the relations among ground states. Quantum phase transitions among these  ground states are explained. In section \ref{rg}, RG equations and obtained phase diagrams are shown. The fully-spin polarized case is briefly discussed in Sec. \ref{strong}. Finally, Sec. \ref{conclusion} is a conclusion and summary.

\section{Model}
\label{model}
\begin{table*}[!tb]
\begin{ruledtabular}
\centering
\begin{tabular}{l|lll}
Case& Fermi momentum & $n$ & symmetries  \\ 
\hline
(a) & $k_{A} = k_{B}$  & $=2$ & U(1)$_{c}$$\times$SU(2)$_{s}$$\times$U(1)$_{o}$$\times$$\mathcal{Z}_{2}$\\
(b) & $k_{A} = k_{B}$  & $\neq 2$&U(1)$_{R}$$\times$U(1)$_{L}$$\times$SU(2)$_{s}$$\times$U(1)$_{o}$$\times$$\mathcal{Z}_{2}$\\
(c) & $k_{A} \neq k_{B}$ & $=2$&U(1)$_{c}$$\times$SU(2)$_{s}$$\times$$\widetilde{\text{U}(1)}_{{o}}$  \\
(d) & $k_{A} \neq k_{B}$  & $\neq 2$ &U(1)$_{R}$$\times$U(1)$_{L}$$\times$SU(2)$_{s}$$\times$$\widetilde{\text{U}(1)}_{{o}}$  \\
(e) & $k_{A} \neq k_{B}=\pi/2$ & $\neq 2$ &U(1)$_{c}$$\times$SU(2)$_{s}$$\times$$\widetilde{\text{U}(1)}_{{o}}$
\end{tabular}
\caption{Possible band structures in the two-orbital Hubbard model, and its symmetries when $v_{A} = v_{B}$. U(1)$_{r}$ represents a gauge transformation of particles with chirality $r$.}
\label{band}
\end{ruledtabular}
\end{table*}
We start from a multi-orbital Hubbard model representing the transition-metal $d$-orbitals with local on-site Coulomb interactions
\begin{equation}
H = \sum_{\langle i , j \rangle } \sum _{m, s} -t^{m m'}_{i j}  \left( c^{\dagger}_{i m s}  c_{j m' s} + \text{h.c.} \right) + H_\text{int}.
\label{kin}
\end{equation}
Here $c^{(\dagger)}_{i m s}$ is the annihilation (creation) operator for a $d$-electron in orbital $m$ with spin $s$ at site $i$. $t_{ij}^{mm'}$ is the hopping between from orbital $m$ on site $i$ to orbital $m'$ on site $j$. The interaction terms $H_{\text{int}}$  will be shown below. We believe this model encapsulates the physics of transition-metal nanowires. Through the paper, we set the  lattice constant to 1. A symmetry breaking field occurs due to the one-dimensional geometry, and the presence of the substrate may further lift the degeneracy of orbitals as well as providing an arbitrary ionization level. Thus, in a general case, one may have many $d$-derived bands with an arbitrary Fermi energy. 

For the sake of simplicity, we will consider here only the cases where two orbitals, $A$ and $B$, are present at the Fermi level. The rotational symmetry in $H_\text{int}$ as we will see always allows us to diagonalize the hopping matrix, so we will ignore off-diagonal terms in the hopping. The band structure is then characterized by four Fermi points: two Fermi momenta, $k_A$ and $k_B$, and two chiralities $r = R, L$, representing electrons around the positive ($R$) and negative ($L$) Fermi momentum. The total particle number is  $n=2(k_A+k_B)/\pi$. In principle there are 5 possible cases which are summarized in Table \ref{band}. In cases (a) and (b), the two Fermi momenta are equal, and the filling is commensurate and incommensurate respectively. In cases (c) and (d), the two Fermi momenta are different, while (c) is at half filling, and (d) is away from half filling. Finally in case (e), one band is commensurate and the other is not, allowing an orbital selective Mott state.

For the two-orbital system the interaction terms have the following form:
\begin{equation}
\begin{split}
H_\text{int} &= U  \sum_{i, m} n_{i m \uparrow} n_{i m \downarrow} \\
&+ U' \sum _{i, s} n_{i A s} n_{i B \overline{s}}\\
&+ (U' -J) \sum _{i, s} n_{i A s} n_{i B s}\\
&- J \sum_{i, s,} c^{\dagger}_{i A s}  c_{i A \overline{s}} c^{\dagger}_{i B \overline{s}} c_{i B s} \\
&+J' \sum_{i, m} c^{\dagger}_{i m \uparrow} c^{\dagger}_{i m \downarrow} c_{i \overline{m} \downarrow} c_{i \overline{m} \uparrow},
\label{int}
\end{split}
\end{equation}
where and $n_{i m s} = c^{\dagger}_{i m s} c_{i m s}$ is the electron density. $\overline{s}=-s$ and $\overline{m}$ means $B(A)$ if $m=A(B)$. Since the substrate will screen the long-ranged Coulomb interaction, we include only on-site Coulomb interactions. $U$ and $U'$ indicates on-site Coulomb repulsion between two electrons in the same band or different bands, and $J$ represents the Hund coupling favoring high spin state. $J'$ is the so called pair-hopping term. For a transition-metal ion in free space all of these parameters are positive.  Now, we assume the following relationship among them, which is usually preserved among $d$-orbitals,
\begin{gather}
J' = J \\
U' = U- 2J.
\label{coulomb integrals}
\end{gather}
To make the symmetry of the interaction terms explicite, we introduce the following charge, spin, and orbital (pseudo-spin) operators:
\begin{gather}
n_{i} = \sum_{ms} n_{ims},\\
\boldsymbol{S}_{i} = \frac{1}{2} \sum_{m s s'} c^{\dagger}_{ims} \boldsymbol{\sigma}_{ss'} c_{ims'},\\
\boldsymbol{T}_{i} = \frac{1}{2} \sum_{m m' s} c^{\dagger}_{ims} \boldsymbol{\tau}_{mm'} c_{im's},
\end{gather}
where $\boldsymbol{\sigma}$ and $\boldsymbol{\tau}$ are Pauli matrices. Then, the interactions in terms of $U$ and $J$ are given by,
\begin{equation}
\begin{split}
H_{\text{int}} &=\sum_{i} \left( \frac{U}{2} n_{i}^2 + J \boldsymbol{S}_i^2 +3J \boldsymbol{T}_i^2 -2J (T^y_{i})^2 - \frac{U+5J}{2}n_i \right).
\label{int2}
\end{split}
\end{equation}
Since $U' >0$, the physically relevant range of $J$ is limited to $0<J<U/2$.

Now we discuss the symmetry of our Hamiltonian at the bare level. When $J=0$, the interaction term possesses U(1)$_{c} \times$SU(4)$_{s, o} \times \mathcal{Z}_{2}$; the indices ``$c$'', ``$s$'', and ``$o$'' denote the charge, spin and orbital parts, and the $\mathcal{Z}_{2}$ symmetry refers to the interchange of orbitals. When $J\neq 0$, the symmetry of the spin-orbital part is broken to SU(2)$_{s} \times$U(1)$_{o}$. We'd like to emphasize that in our convention, the U(1) axis is the orbital $y$-axis, wheres it is orbital $z$-axis in the convention of Refs. \onlinecite{Nonne2010} and  \onlinecite{Lee2004}. Only if $v_{A} = v_{B}$ and $k_{A}=k_{B}$, as in cases (a) and (b), does the total Hamiltonian have the same symmetry as the interaction. Otherwise, including cases (c)--(e), the total symmetry is reduced to U(1)$_{c} \times$SU(2)$_{s}$ at the bare level due to the lower symmetry of the kinetic term. However, it is known that low-energy theories in weak-coupling still have an effective $\widetilde{\text{U(1)}_{o}}$ symmetry\cite{Lin1998,Controzzi2005,Boulat2009,Nonne2010} at least if $v_{A} = v_{B}$. We will see this in detail later.

\section{Bosonization}
\label{bosonization}
\begin{figure}[!tb]
\centering
\includegraphics[scale=0.52]{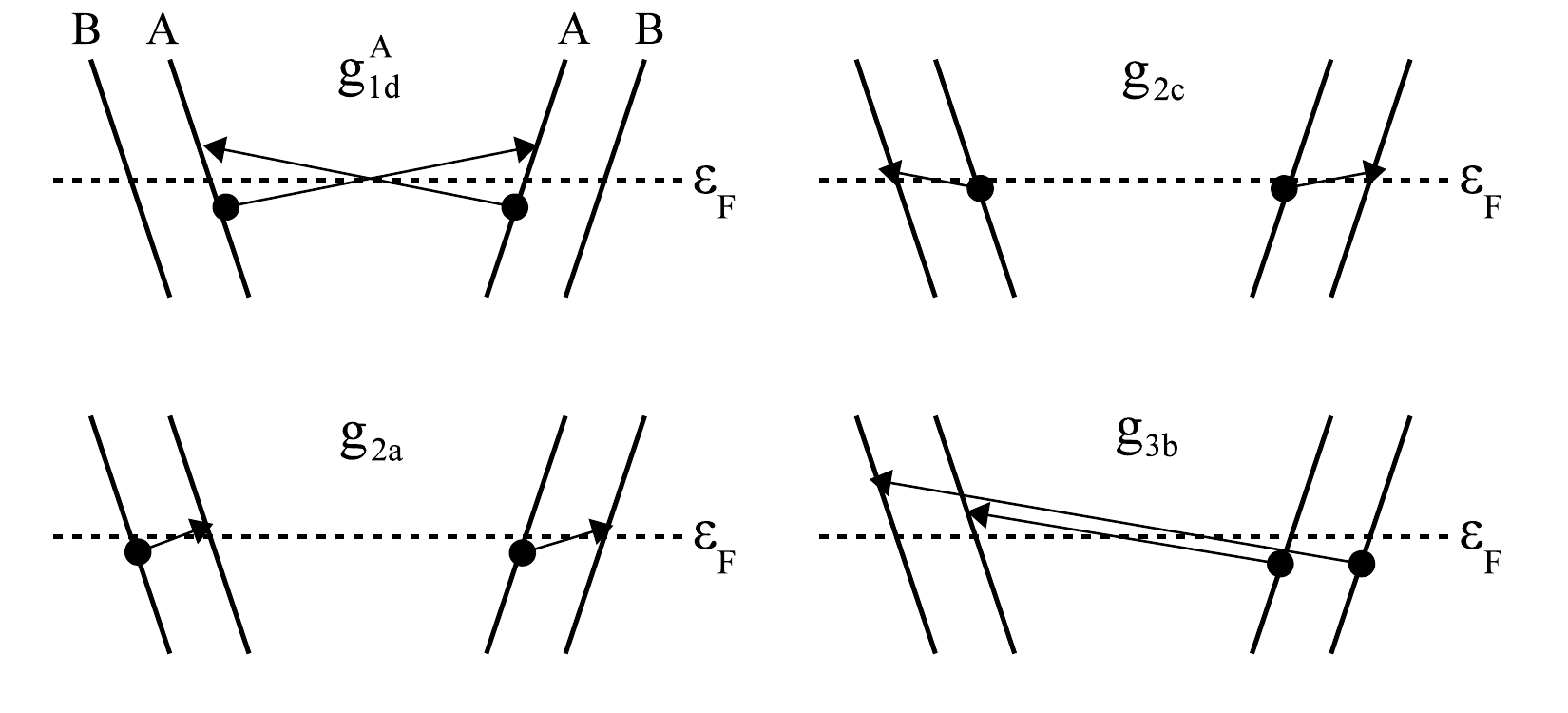}
\caption{Various scattering processes and ``g-ology"}
\label{gology}
\end{figure}
In 1D systems, it is known that bosonization enables us to describe the low energy physics in a  simple manner. In this section we present an Abelian-bosonization analysis, which is useful for the cases when difference in velocities is negligible. Appendix \ref{bf} outlines the more complicated formalism needed for unequal velocities.

To classify the various scattering processes we follow the notation of Ref. \onlinecite{Chudzinski2008}. ``1--4" corresponds to conventional ``g-ology" indices for left and right moving fields, and ``$a$--$d$" are similar but label orbital indices. Some examples are given in Fig. \ref{gology}. The SU(2)$_{s}$ symmetry constrains coupling constants as 
\begin{equation}
\begin{split}
g^{m}_{1d} -g^{m}_{2d}=0\\
g^{\perp}_{2b} - g^{\parallel}_{2b} - g_{1a}=0\\
g_{1c} -g_{2c}+g_{\parallel c}= 0.
\end{split}
\end{equation}
For the rest of the paper, we will omit $\perp$ when it is not confusing.

When two Fermi points coincide, we have additional processes $g_{\parallel a}$, $g_{1b}$, and $g_{2a}$, which are connected by the SU(2)$_{{s}}$ symmetry as,
\begin{equation}
g_{\parallel a} - g_{1b} + g_{2a} =0.
\end{equation} 
$k_{A} = k_{B}$ means that we have the U(1)$_{{o}}$ symmetry, and this implies
\begin{equation}
\begin{split}
g^{A}_{1(2)d}-g^{B}_{1(2)d}=0\\
-g_{2d}^{m}+g_{2b}^{\perp}+g_{2c}+g_{2a}=0\\
-g_{1d}^{m}+g_{1b}+g_{1c}+g_{1a}=0\\
g^{\parallel}_{2b}+g_{\parallel c} -g_{\parallel a}=0.
\end{split}
\end{equation}

As Umklapp processes, we have $g_{\parallel b}$ and $g_{3i}$ ($i = a$--$d$). The $g_{3d}^{m}$ process represents the intraband Umklapp process, so it only exists for $k_{m}= \pi/2$. Other Umklapp processes are possible whenever $n=2$. The SU(2)$_{{s}}$ symmetry gives
\begin{equation}
g_{3a}+g_{\parallel b} -g_{3b}=0,
\end{equation}
and the U(1)$_{{o}}$ symmetry leaves
\begin{equation}
g_{3d}=g_{3a}+g_{3b}+g_{3c}.
\end{equation}

Using standard bosonization,\cite{Voit1995a,giamarchi2004quantum, gogolin2004bosonization} the Hamiltonian density $\mathcal{H}_{0}$ of the free-boson part becomes,
\begin{equation}
\mathcal{H}_{0}= \frac{1}{2\pi}\sum_{\substack{\mu = c, s\\ \nu = 0, \pi}} v_{\mu \nu} \left[ K_{\mu \nu} ( \nabla \theta_{\mu \nu} )^2 + \frac{1}{K_{\mu \nu} }( \nabla \phi_{\mu \nu} )^2 \right]
\end{equation}
where $\phi$ and $\theta$ are connected to density and current: $\nabla \phi \propto n$, and $\nabla \theta \propto j$. The renormalized Luttinger parameters and velocities are given by
\begin{equation}
\begin{split}
K_{c 0(\pi)} &= \sqrt{\frac{1 - (y_{1d} \pm y_{2b})/2}{1+(y_{1d} \pm y_{2b})/2}} \equiv 1-y_{c0(\pi)}\\
K_{s 0(\pi)} &= \sqrt{\frac{1+ (y_{1d} \pm y_{1a})/2}{1- (y_{1d} \pm y_{1a})/2}} \equiv 1-y_{s0(\pi)}\\
v_{c 0(\pi)} &=  v\sqrt{1-(y_{1d} \pm y_{2b})^{2}/4}\\
v_{s 0(\pi)} &=  v\sqrt{1-(y_{1d} \mp y_{1a})^{2}/4}
\end{split}
\end{equation}
with $y_{i} \equiv g_{i}/4\pi v$ and $y_{2b} \equiv y_{2b}^{\perp} + y_{2b}^{\parallel}$. ``$c$''  and ``$s$''  represent the charge and spin modes, and ``$0$''  and ``$\pi$''  are used for bonding and antibonding combination. We introduced $y_{\mu \nu}$ for each Luttinger parameter for later use. A detailed derivation of this Hamiltonian is given in Appendix \ref{bos}. We ignore the velocity difference induced by the $g_{4}$ process, since its effect is to shift the phase boundaries slightly. We will provide a separate treatment for systems in which difference of initial velocities is quite large.

The SU(2)$_{s}$ symmetry fixes $K_{s0(\pi)}$ to be
\begin{equation}
K_{s 0(\pi)} = 1+ (y_{1d}\pm y_{1a})/2 \equiv1 -y_{s0(\pi)}
\end{equation}
along the RG flow. Similarly the U(1)$_{o}$ symmetry (when it exists) constrains $K_{c\pi}$ to be
\begin{equation}
K_{c\pi} = 1 + (- y_{1d} +y_{2b})/2\equiv1 -y_{c\pi}.
\end{equation}

The interaction part of the Hamiltonian is rather complicated. The interaction terms common to all cases are
\begin{equation}
\begin{split}
H_\text{int} 
&= g_{1d} \int  \cos \left( 2 \phi_{s0} \right)    \cos \left( 2 \phi_{s\pi} \right)\\
&+ g_{1a} \int  \cos \left( 2 \phi_{s0} \right)    \cos \left( 2 \theta_{s\pi} \right) \\
&-  g_{1c} \int  \cos \left( 2 \phi_{s0} \right)    \cos \left( 2 \theta_{c\pi} \right) \\
&-  g_{2c} \int  \cos \left( 2 \phi_{s\pi } \right) \cos \left( 2 \theta_{c\pi} \right) \\ 
&+ g_{\parallel c} \int  \cos \left( 2 \theta_{s\pi } \right) \cos \left( 2 \theta_{c\pi} \right).
\end{split}
\end{equation}
Here $\int$ stands for $\frac{ 1}{(2\pi \alpha)^2}\int dx $, and $\alpha$ is the cut-off which is of the order of the lattice constant.  The last term does not exist in the original Hamiltonian, but will be generated after renormalization.

When $k_{A} = k_{B}$, in case (a) and (b), additional processes are allowed,
\begin{equation}
\begin{split}
H'_\text{int} 
&= g_{\parallel a} \int   \cos \left( 2 \phi_{s\pi} \right)\cos \left( 2 \phi_{c\pi}  \right)\\
&+ g_{1b}  \int   \cos \left( 2 \phi_{s0} \right)   \cos \left( 2 \phi_{c\pi}\right)\\
&+  g_{2a}  \int   \cos \left( 2 \theta_{s\pi} \right)\cos \left( 2 \phi_{c\pi} \right).
\end{split}
\end{equation}

Finally when the filling is commensurate, we have
\begin{equation}
\begin{split}
H''_\text{int} 
&= g_{3a} \int   \cos \left( 2 \phi_{c0} \right)\cos \left( 2\theta_{s\pi} \right)\\
&- g_{3b}  \int   \cos \left( 2 \phi_{c0} \right)\cos \left( 2 \phi_{s\pi} \right)\\
&- g_{3c}  \int   \cos \left( 2 \phi_{c0} \right)\cos \left( 2 \theta_{c\pi} \right)\\
&- g_{3d}  \int   \cos \left( 2 \phi_{c0} \right)\cos \left( 2 \phi_{c\pi} \right)\\
&- g_{\parallel b} \int  \cos \left( 2 \phi_{c0} \right)\cos \left( 2 \phi_{s0} \right).
\end{split}
\end{equation}
The $g_{3d}$ process exists only when both bands are commensurate, case (a). Again, we ignored the $g_{4}$ type interactions whose scaling dimension is always larger than 2, and this is consistent with the equal velocity approximation. The initial values of these coupling constants are: $g_{1d} = g_{3d}= 4U$, $g_{1a} = g_{1c} =g_{2c} = g_{2a} = g_{3a} = g_{3c} = 4J$, $ g^{\parallel}_{2b}= g_{\parallel a} = g_{\parallel b} = 4(U-3J)$, and $g_{1b} = g^{\perp}_{2b} =g_{3b} = 4(U-2J)$. They will take different values after renormalization.

Finally we will discuss the symmetry of the linearized model (Table \ref{band}). First, U(1)$_{c}$ and SU(2)$_{s}$ (around $z$-axis) are displayed in the invariance of the Hamiltonian under the translation of $\theta_{c0}$ and $\theta_{s0}$. In fermionic language, each corresponds to the following gauge transformation:
\begin{equation}
c_{rms} \rightarrow e^{i \alpha }c_{rms},  \ \ c_{rms} \rightarrow e^{i s \alpha }c_{rms}.
\end{equation}
The indices $r$, $m$, and $s$ represent chirality, orbital, and spin, and $\alpha$ expresses the constant phase shift. The conserved Noether currents, corresponding to the U(1)$_{c}$ and SU(2)$_{s}$ symmetries are,
\begin{equation}
\begin{split} 
\sum_{r m s} N_{rms} &\propto \int dx \nabla \phi_{c0}\\
\sum_{r m s} s N_{rms} &\propto \int dx \nabla \phi_{s0},
\end{split}
\end{equation}
where $N_{rms}$ is the particle number at branch specified by $r$, $m$, and $s$. Since $\nabla \phi$ is momentum conjugate of $\theta$, the operator $\exp \left( \int dx \nabla \phi \right)$ gives a constant shift of $\theta$. The SU(2)$_{s}$ rotation around $x$- and $y$-axis are not manifest in Abelian bosonization.

Away from half filling, there is also a continuous chiral symmetry under the transformation, $c_{rms} \rightarrow e^{i r \alpha }c_{rms}$. Thus, the Hamiltonian is invariant under arbitrary translation of $\phi_{c0}$, with conserved total currents, $J_{A}+J_{B} \propto \int dx \nabla \theta_{c0}$ where $J_{m} = \sum_{s} N_{Rms}-N_{Lms}$. At half filling, this symmetry is broken to a discrete symmetry, and true long-range order can be realized.\cite{Wu2003} When a system has both the chiral symmetry and the U(1)$_{c}$ symmetry, this implies that left, and right moving parts have separate conservation laws corresponding to the U(1)$_{R}$$\times$U(1)$_{L}$ symmetry.  Similarly, other gauge transformations such as 
\begin{equation}
\begin{split}
c_{rms} &\rightarrow e^{i m  \alpha }c_{rms}\\
c_{rms} &\rightarrow e^{i m s \alpha }c_{rms}\\
c_{rms} &\rightarrow e^{i r s\alpha }c_{rms} \\
c_{rms} &\rightarrow e^{i r m s\alpha }c_{rms}
\end{split}
\end{equation}
leave the Hamiltonian invariant for discrete values of  $\alpha = n \pi/2 $; each corresponds to the discrete shift of $\theta_{c\pi}$, $\theta_{s\pi}$, $\phi_{s0}$, and $\phi_{s\pi}$. 

The effective $\widetilde{\text{U}(1)}_{{o}}$ symmetry appearing when $k_{A}\neq k_{B}$ and $v_{A} = v_{B}$,\cite{Lin1998,Controzzi2005,Boulat2009,Nonne2010} corresponds to the invariance under the translation of $\phi_{c\pi}$ or the gauge transformation,
\begin{equation}
c_{rms} \rightarrow e^{i r m \alpha }c_{rms} .
\end{equation}
The conserved ``charge'' corresponding to this symmetry is the difference of two orbital currents: $J_{A} - J_{B} \propto \int dx \nabla \theta_{c\pi}$. 

When  $k_A =k_B$, there is an explicit orbital rotational symmetry about $y$-axis. This transformation mixes fermions in different orbitals, so its generator cannot be expressed as a local operator in Abelian bosonization. This fact leads to a new combination of possible ground states as will be shown.

\section{Order parameters}
\label{order parameters}
\subsection{Order parameters away from half filling}
As order parameters, we take fermion bilinears characterized by the chirality, spin and orbital indices; thus in the model considered here there are particle-hole bilinears,
\begin{equation}
(\Delta_\text{ph})_{rr'}^{ss';mm'}=  c^\dagger_{rms}c_{r'm's'} 
\end{equation}
and particle-particle bilinears,
\begin{equation}
(\Delta_\text{pp})_{rr'}^{ss';mm'}=  m s c^\dagger_{rms}c^{\dagger}_{r'\overline{m'} \overline{s'}},
\end{equation}
where $c^{(\dagger)}_{rms}$ is the annihilation (creation) operator of electron with chirality $r$, orbital $m$, and spin $s$. Since combinations of $r = r'$ are irrelevant in a RG sense, we will fix $(r, r') = (R, L)$ in the remainder of this paper. We will use the following convenient basis to represent them,
\begin{gather}
\mathcal{O}^{ij}_\text{ph} =\sum_{mm'ss'} \tau^{i}_{mm'} \sigma^{j}_{ss'} (\Delta_\text{ph})^{ss';mm'} + \text{h.c.}\\
\mathcal{O}^{ij}_\text{pp} = \sum_{mm'ss'} \tau^{i}_{mm'} \sigma^{j}_{ss'} (\Delta_\text{pp})^{ss';mm'} + \text{h.c.}
\end{gather}
where $i, j = (0, 1, 2, 3)$ and $\tau$ and $\sigma$ are Pauli matrices with $\tau^{0}_{ab} = \sigma^{0}_{ab} = \delta_{ab}$. These transform as rank 2 tensors under SO(4) $\simeq$ SU(2)$_{{s}}\times $SU(2)$_{{o}}$ transformations; the SU(2)$_{{s}}$ rotations connect $\sigma^{1,2,3}$, and the U(1)$_{{o}}$ rotation (if it exists) connects $\tau^{1}$ and $\tau^{3}$. Thus, we take the quantization axis along $z$-direction for spins. Additionally, as we will show below, all the high spin states ($j=3$) such as SDW states and triplet superconductivities are excluded from the possible ground states in our models, so we will not consider them.

We label our order parameters by the transferred momentum and the sign at each Fermi point. Here we have 4 Fermi points each degenerate about spins, so, in principle, there are 4 possible cases (Table \ref{angular}). We use $s$-wave when all 4 points have the same signs. $p_{x}$ and $p_{y}$ are odd under the inversion $R\leftrightarrow L$, and $A \leftrightarrow B$ respectively. $d$-wave is odd under both inversions.

\begin{table}[!b]
\begin{ruledtabular}
\centering
\begin{tabular}{c|cccc}
``Angular momentum''& $(A, R)$ & $(A, L)$ & $(B, R)$ & $(B, L)$\\
\hline
$s$ & $+$ & $+$ & $+$ & $+$ \\
$p_{x}$ & $+$ &$-$ & $+$ &$-$\\
$p_{y}$ & $+$ & $+$ & $-$ &$-$\\
$d$ & $+$  & $-$ & $-$ & $+$ 
\end{tabular}
\caption{Angular momentum and sign distributions}
\label{angular}
\end{ruledtabular}
\end{table}

Applying this classification, we find that $i=0, 1$ are both $s$-wave for the particle-hole channel, while the former is intraband type and the latter is interband type. We put `` ' '' for an interband order to distinguish these two. $i =2$ is found to be interband $p_{y}$-wave, and $i=3$ is intraband $p_{y}$-wave. For the particle-particle channel, we found $d'$-, $p_{y}$-, $s$-, and $s'$-wave orders for $i = 0, \ldots, 3$ accordingly. Since $p_{x}$-wave does not appear, we will use $p$ for $p_{y}$-wave orders in the rest of the paper. We note that the $d$-wave superconductivity, which often appears in two-leg ladder models,\cite{Schulz1996} is the $p$SS state in our notation. The result is summarized in Table \ref{order}.

The order parameters in bosonized forms are,
\begin{equation}
\begin{split}
\mathcal{O}^{00}_\text{ph} &\propto \cos \left(k_0 x-\phi _{c0}\right) \sin \left(k_{\pi } x-\phi _{c\pi}\right) \sin \left(\phi _{s0}\right) \sin \left(\phi _{s\pi}\right)\\
\mathcal{O}^{10}_\text{ph} &\propto \cos \left(k_0 x-\phi _{c0}\right) \sin \left(\theta _{c\pi}\right) \cos \left(\phi _{s0}\right) \cos \left(\theta _{s\pi}\right)\\
\mathcal{O}^{20}_\text{ph} &\propto \cos \left(k_0 x-\phi _{c0}\right) \cos \left(\theta _{c\pi}\right) \cos \left(\phi _{s0}\right) \cos \left(\theta _{s\pi}\right)\\
\mathcal{O}^{30}_\text{ph} &\propto \cos \left(k_0 x-\phi _{c0}\right) \sin \left(k_{\pi } x-\phi _{c\pi}\right) \cos \left(\phi _{s0}\right) \cos \left(\phi _{s\pi}\right)
\end{split}
\end{equation}
\begin{equation}
\begin{split}
\mathcal{O}^{00}_\text{pp} &\propto e^{-i \theta _{c0}} \sin \left(k_{\pi} x-\phi _{c\pi}\right) \cos \left(\phi _{s0}\right) \cos \left(\theta _{s\pi}\right)\\
\mathcal{O}^{10}_\text{pp} &\propto e^{-i\theta _{c0}} \sin  \left( \theta _{c\pi }\right)  \sin  \left( \phi _{s0}\right)  \sin  \left( \phi _{s\pi }\right)\\
\mathcal{O}^{20}_\text{pp} &\propto e^{-i \theta _{c0 }} \cos \left(\theta _{c\pi }\right) \sin \left(\phi _{s0 }\right) \sin \left(\phi _{s\pi }\right)\\
\mathcal{O}^{30}_\text{pp} &\propto e^{-i \theta _{c0}} \sin \left(k_{\pi} x-\phi _{c\pi}\right) \sin \left(\phi _{s0}\right) \sin \left(\theta _{s\pi}\right),
\end{split}
\end{equation}
with $k_{0(\pi)} = k_{A} \pm k_{B}$. If coupling constants grow to the order of $t$ after renormalization, the corresponding bosonic fields are pinned to the values which minimize the resultant potential. For incommensurate fillings, the total charge mode is massless, and the interaction terms pin the other modes to definite values (in mod of $\pi/2$). The pinned values determine the order parameter which gives a finite value of correlation. It is easy to find such pinned values from the above expressions e.g. $\left( \theta_{c\pi}, \phi_{s0},\phi_{s\pi}\right)  = \left(\pi/2,0,0 \right)$ for the $s'$CDW state.

\begin{table*}[!htb]
\begin{ruledtabular}
\centering
\begin{tabular}{llll}
$(i,j)$&Types of particle-hole order & Types of particle-particle order \\ 
\hline
$(0,0)$ & Charge density wave (CDW) & $d'$-wave singlet SC($d'$SS)  \\
$(1,0)$& $s'$-wave charge density wave ($s'$CDW)  & $p_{y}$-wave singlet SC ($p$SS) \\$(2,0)$ & $p'_{y}$-wave charge density wave ($p'$CDW)  &  $s$-wave singlet SC ($s$SS)  \\
$(3,0)$ & $p_{y}$-wave charge density wave ($p$CDW) &  $s'$-wave singlet SC ($s'$SS)
\end{tabular}
\caption{Classification of order parameters. `` ' '' indicates that the order is inter-orbital type. We use $p$ to express $p_{y}$-wave orders, since $p_{x}$-wave orders do not appear.}
\label{order}
\end{ruledtabular}
\end{table*}

\subsection{Order parameters at half filling}
When a band is commensurate, the number of possible ground states increases since the total charge mode also comes into play. After renormalization, all modes become massive unless the system remains in a Luttinger Liquid phase. The phases are classified by the fixed values of pinned fields.

The first set of new order parameters we consider is ``bond" orders, which is basically the density-wave slid from on-site to ``on-bond". Away from half filling, site and bond density waves are degenerate and the total charge mode is massless. However, at half filling, these two states decouple, and their expectation values of $\langle \phi_{c0} \rangle $ are different by $\pi/2$. We call these on-bond states BDW (bond density wave). Each phase is doubly degenerate about the translation of $\langle \phi_{c0} \rangle $ by $\pi$.

A second class of insulating phases evolves into superconducting phases upon doping. These insulating phases share similar properties with the superconducting phases in the sense that the pinned fields and values are the same except for the total charge mode. While we will explain each state in detail later, we show the results first. At half filling, the $s'$SS state becomes the Haldane charge (HC)\cite{Nonne2010} state or the rung-singlet (RS) state.\cite{Nishiyama1995} Similarly, the $d'$SS state turns into the Haldane-orbital (HO) state\cite{Nonne2010} or the rung-triplet (RT) state.\cite{Nishiyama1995} The S-Mott and S'-Mott phases are developed from the $s$SS state, and the D-Mott and D'-Mott phases are from the $p$SS state. Two insulating phases developed from the same superconducting state are connected by the shift of $\phi_{c0}$ by $\pi/2$.

Another way of looking at these insulating phases is to regard them as Ising disorder phases of the CDW and BDW phases.\cite{Tsuchiizu2002} The disordered states can be obtained by applying the mapping, $\phi_{s\pi} \leftrightarrow \theta_{s\pi}$, to density-wave states. While the order parameters of density-wave states can be expressed by the local fermions, these disordered phases need non-local string operators. The Ising dual phases of the $s'$CDW(BDW) state is the S(')-Mott state, and the $p'$CDW(BDW) state is dual to the D(')-Mott state. The RS, RT, HC, and HO states are Ising dual to the BDW, $p$BDW, CDW, and $p$CDW states respectively. These disordered states are non-degenerate unlike density-wave ground states. The correspondence between insulating/metallic phases and ordered/disordered phases is summarized in Table \ref{doping}. The pinned values of 16 insulating phases can be found in Ref. \onlinecite{Nonne2010}. 

\begin{table}[!b]
\begin{ruledtabular}
\centering
\begin{tabular}{cc|cc}
\multicolumn{2}{c|}{Ordered }&\multicolumn{2}{c}{Disrdered}\\
\hline
Insulating & Metallic  &Insulating   & Metallic \\
\hline
CDW & \multirow{2}{*}{CDW}& HC & \multirow{2}{*}{$s'$SS}\\
BDW(SP) &&RS &\\ 
\hline
$s'$CDW&  \multirow{2}{*}{$s'$CDW}&S-Mott &\multirow{2}{*}{$s$SS}\\
$s'$BDW & & S'-Mott &\\
\hline
$p'$CDW(SF) &   \multirow{2}{*}{$p'$CDW}&D-Mott &  \multirow{2}{*}{$p$SS}\\
$p'$BDW (FDW) &&D'-Mott &\\
\hline
$p$CDW & \multirow{2}{*}{$p$CDW}&HO &\multirow{2}{*}{$d'$SS}\\
$p$BDW(SP$_{\pi}$) & &RT &
\end{tabular}
\caption{Correspondence between insulating/metallic phases and ordered/disordered phases. The commonly used names are given in parenthesis. SP: spin-Peierls, SF: staggered flux, FDW: $f$-wave density wave.}
\label{doping}
\end{ruledtabular}
\end{table}

To see the above properties of each phase more carefully, we consider two-particle states corresponding to each superconducting state. We first look at the $s$SS and $p$SS states. Considering the occupation of orbital $A$ and $B$ at site $i$, we find
\begin{equation}
\begin{split}
\Ket{s\text{SS}} &=   \prod_{i} \left[ c_{i A\uparrow}c_{ iA\downarrow} + c_{i B\uparrow}c_{ iB\downarrow} \right]\Ket{0}\\
\Ket{p\text{SS}} &=   \prod_{i} \left[ c_{i A\uparrow}c_{ iA\downarrow} - c_{i B\uparrow}c_{ iB\downarrow} \right] \Ket{0}.
\end{split}
\end{equation}
These two-particle states are the same as the ones given in Ref. \onlinecite{Tsuchiizu2002} as the S-Mott and D-Mott states in the bonding/antibonding basis. Therefore, we identify the corresponding insulating states as the S-Mott and D-Mott states. The S'-Mott and D'-Mott states are different from these only by $\delta \phi_{c0} = \pi/2$, and this indicates that the order is located on bond. To be precise, while the authors of Ref. \onlinecite{Tsuchiizu2002} claimed that the D-Mott and S'-Mott states are connected to the RS and RT states in the strong coupling limit, this is not the case in our models as we will see later. The discrepancy comes from the fact that the strong transverse hopping requires the use of bonding/antibonding basis in weak coupling analysis, although the strong coupling analysis starts from two independent ladders. However, in the weak coupling regime, the properties of Mott states are the same.

\begin{figure}[!tb]
\centering
\subfigure[rung-singlet (RS)]{
\includegraphics[scale=0.8]{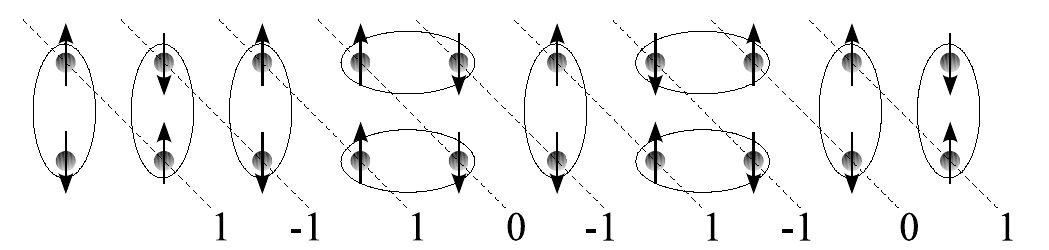}
\label{rs}}
\subfigure[rung-triplet (RT)]{
\includegraphics[scale=0.8]{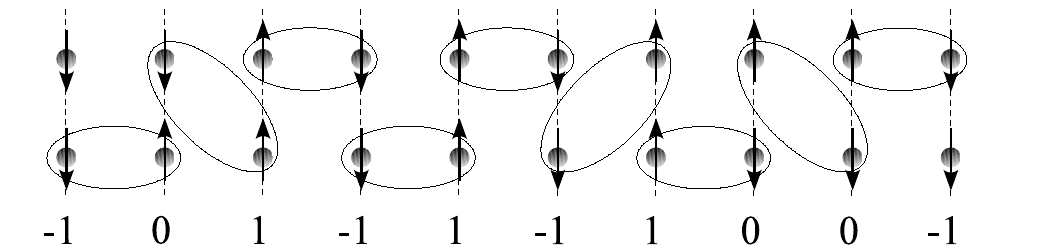}
\label{rt}}
\caption{Rung singlet and rung triplet states (after Ref. \onlinecite{Nishiyama1995}). Closed circles represent singlet bonds, and dashed lines indicate hidden antiferromagnetic order.}
\label{rvb}
\end{figure}

Next, the $s'$SS and $d'$SS states can be represented by the following two particle states, 
\begin{equation}
\begin{split}
\Ket{s'\text{SS}}&= \prod_{i} \left[ c_{i A\uparrow}c_{ i B\downarrow} - c_{i A\downarrow}c_{ iB\uparrow} \right]\Ket{0}\\
\Ket{d'\text{SS}}&= \prod_{i}   \left[ c_{i A\uparrow}c_{ i+1, B\downarrow} - c_{i A\downarrow}c_{i+1, B\uparrow} -(A\leftrightarrow B) \right]\Ket{0},
\label{twoparticle}
\end{split}
\end{equation}
where we ignored the total charge mode, and assumed $k_{A} =k_{B} =\pi/2$ for the $d'$SS state. Obviously, the $s'$SS state forms a singlet Cooper pair on a rung, while the $d'$SS state forms such singlet pair in staggered way. These pairings are the reminiscent of resonance-valence-bond (RVB) in the RS and RT states which appear in spin-$\frac{1}{2}$ Heisenberg two-leg ladders.\cite{Nishiyama1995, Kim2000} The RS state appears in chains with antiferromagnetic coupling along rung (or ferromagnetic coupling over plaquette diagonals), and it is a RVB state whose stable configurations are the singlet along rung or ladder as in Fig. \ref{rs}. We can see the clear connection between the RS state and the $s'$SS states in Eq. \eqref{twoparticle}. This state is characterized by a non-zero expectation value of a string operator
\begin{equation}
\langle \left(S_{A,i}^{z}+S_{B,i+1}^{z} \right) e^{i\pi\sum_{k=1}^{j-1}  \left(S_{A,k}^{z}+S_{B,k+1}^{z} \right)}  \left(S_{A,j}^{z}+S_{B,j+1}^{z} \right)\rangle,
\label{string}
\end{equation}
and exhibits a ``hidden antiferromagnetic order''; as we group $S_{A,i}$ and $S_{B,i+1}$, the total spin align antiferromagnetically along ladder except spin-0 site.

For chains coupled ferromagnetically along rung (or antiferromagnetically over plaquette diagonals), the above order disappears, and the Valence-Bond-Solid (VBS) configuration\cite{Affleck1987} becomes stable as in Fig. \ref{rt}. As can be seen, singlet pairs are formed in a staggered manner as in the $d'$SS state, and this results in triplet pair along a rung. Thus, we call it rung-triplet (RT). The following order parameter takes a non-vanishing value,
\begin{equation}
\langle \left(S_{A,i}^{z}+S_{B,i}^{z} \right) e^{i\pi\sum_{k=1}^{j-1}  \left(S_{A,k}^{z}+S_{B,k}^{z} \right)}  \left(S_{A,j}^{z}+S_{B,j}^{z} \right)\rangle,
\label{haldane}
\end{equation}
which represents a hidden order about the spin-triplet on a rung [Fig. \ref{rt}]. This state smoothly continues to the Haldane gapped state as interchain coupling becomes large.

The string operators for these disordered states are non-local, so complications arise in the transcription to bosonic variables. A generally accepted form for Eqs. \eqref{string} and \eqref{haldane} is\cite{Kim2000, Berg2008}
\begin{equation}
\langle \sin(\phi_{s0}(x))\sin(\phi_{s0}(y))\rangle.
\end{equation}
This correlation function takes a non-zero value also for the $s'$SS and $d'$SS phases. Considering these properties, we conclude that the RS and RT states are the corresponding insulating states of the $s'$SS and $d'$SS states. In particular, we found that the $d'$SS and RT states are the disordered versions of high spin states. As we mentioned, the D-Mott and S'-Mott states do not evolve to the RS and RT states.

The bonding counterparts of the RS and RT phases are the Haldane-charge (HC), and Haldane-orbital (HO) phases proposed by Nonne \textit{et al.}\cite{Nonne2010} as the Haldane gapped states of pseudo spin-1 antiferromagnetic Heisenberg chain; this realizes when the charge or orbital symmetry is promoted from U(1) to SU(2). The form of string operators for these states is similar to Eqs. \eqref{string} and \eqref{haldane} if we replace SU(2) spin operators by the charge and orbital SU(2) operators,
\begin{align}
J^{z}_{i} =\frac{1}{2}( n_{A,i}+n_{B,i}), & \ \  J^{\dagger}_{i} =c^{\dagger}_{i,A\uparrow}c^{\dagger}_{i,B\downarrow}-c^{\dagger}_{i,A\downarrow}c^{\dagger}_{i,B\uparrow}\\
T^{z}_{i} = n_{A,i} -n_{B,i}, & \ \ T^{\dagger}_{i} =c^{\dagger}_{i,A\uparrow}c_{i,B\uparrow}+c^{\dagger}_{i,A\downarrow}c_{i,B\downarrow}.
\label{orbital}
\end{align}
The bosonized form of string operators becomes 
\begin{gather}
\langle \sin(\phi_{c0}(x))\sin(\phi_{c0}(y))\rangle \\
\langle \sin(\phi_{c\pi}(x))\sin(\phi_{c\pi}(y))\rangle,
\end{gather}
where $\nabla \phi_{c0} \sim J^{z}(x)$ and $\nabla \phi_{c\pi} \sim T^{z}(x)$. These expressions agree with the fact that in the insulating phase the total charge mode should be massive, and that $\phi_{c\pi}$ is fixed also in the $s'$SS and $d'$SS states. Further study of the connection between phases appearing in strong-coupling limit, and those in the weak coupling limit would be of interest, but will not be pursued here, because our focus is on the weak coupling limit.

\section{Duality, and dynamical symmetry enlargement}
\label{duality}
\subsection{Possible ground states}

\begin{table*}[!thb]
\begin{ruledtabular}
\centering
\begin{tabular}{c|cccccccc}
&$\mathbf{1}$ & $\Omega_{c}$&$\Omega_{o}$ &$\Omega_{c,o}$ & $\Omega_{I}$& $\Omega_{c,I}$ & $\Omega_{s}$ &$\Omega_{c,s}$\\
\hline
 (a) &BDW & CDW &  $p'$CDW  &$p'$BDW  & S-Mott &S'-Mott &RT & HO\\
 (b) & CDW & CDW & $p'$CDW & $p'$CDW& $s$SS& $s$SS& $d'$SS & $d'$SS\\
 (c) &S-Mott &S'-Mott & D-Mott & D'-Mott & $s'$CDW  & $s'$BDW & $p'$BDW & $p'$CDW\\
 (d) & $s$SS &  $s$SS &$p$SS &$p$SS & $s'$CDW & $s'$CDW & $p'$CDW & $p'$CDW
\end{tabular}
\caption{Possible ground states for case $(a)$--$(d)$ in Table \ref{band} at equal velocities. The BDW, CDW, S-Mott, and $s$SS phases in the second column are expressed by the fundamental SO(8) or SO(6) Gross-Neveu model for case $(a)$--$(d)$ respectively. The other states are mapped from these fundamental states by the duality transformation $\Omega$ in the top row. $\Omega_{\nu}$ is an operation of $\xi^{a}_{L} \rightarrow - \xi^{a}_{r}$ for all the $a$'s in a symmetry sector $\nu$. For example, in case (a), the BDW state is mapped to the $p'$CDW state by $\Omega_{o}$. For doped cases, (b) and (d), the charge mode is separated from the rest, and the ground states are invariant under $\Omega_{c}$.}
\label{ground states}
\end{ruledtabular}
\end{table*}

Two of the most prominent features of one-dimensional systems are \textit{duality}\cite{Boulat2009,Nonne2010} and \textit{dynamical symmetry enlargement (DSE)}.\cite{Lin1998, Konik2002} The idea of duality is based on the observation that the low energy theory is invariant under some discrete operations apart from the continuous symmetries listed in Table \ref{band}. These discrete symmetries enable us to relate one ground state to another, and to understand quantum phase transitions among them. 

DSE means that the effective theory describing the low energy fixed point exhibits a higher symmetry than that of the original lattice Hamiltonian. This phenomenon was noted by Lin \textit{et al.}\cite{Lin1998}, who found that the low-energy theory of half-filled two-leg Hubbard ladder is the SO(8) Gross-Neveu (GN) model. Since their work, DSE has been seen in other multiband systems.\cite{Assaraf2004, Bunder2007} Combining these two ideas, we will identify the possible ground states for our models, taking DSE for granted.

Now, as a preparation, in order to exhibit the symmetries of the Hamiltonian, we refermionize the model using 8 Majorana fermions as explained in Refs. \onlinecite{Nonne2010}, and \onlinecite{Tsvelik2011}. We decompose each mode into two Majorana fermions as
\begin{equation}
\begin{split}
\psi^{c0}_{r} =\frac{1}{\sqrt{2}}\left( \xi^{7}_{r} + i \xi^{8}_{r}\right), & \ \ \psi^{c\pi}_{r} =\frac{1}{\sqrt{2}}\left( \xi^{5}_{r} + i \xi^{6}_{r}\right)\\
\psi^{s0}_{r} =\frac{1}{\sqrt{2}}\left( \xi^{1}_{r} + i \xi^{2}_{r}\right), & \ \ \psi^{s\pi}_{r} =\frac{1}{\sqrt{2}}\left( \xi^{4}_{r} + i \xi^{3}_{r}\right).
\end{split}
\label{majorana}
\end{equation}
At half filling, the obtained expression for the U(1)$_{o}$ symmetric case, (a), is,
\begin{equation}
\begin{split}
\mathcal{H} &= -i\frac{v}{2\pi} \sum_{a=1}^{8}\left( \xi^{a}_{R}\partial\xi^{a}_{R}- \xi^{a}_{L}\partial\xi^{a}_{L}   \right) \\
&+ \frac{g_{1}}{2} \kappa_{s}^{2} + g_{2} \kappa_{s}\kappa_{o} +g_{3}\kappa_{s}\kappa_{I} \\
&+ g_{4}\kappa_{o}\kappa_{I} +\frac{g_{5}}{2} \kappa_{o}^{2} +\frac{g_{6}}{2}\kappa_{c}^{2} \\
&+ g_{7}\kappa_{s}\kappa_{c}+ g_{8}\kappa_{o}\kappa_{c}+ g_{9}\kappa_{I}\kappa_{c}
\end{split}
\label{so8}
\end{equation}
where $\kappa_{s} = \sum_{a=1}^{3} \xi^{a}_{R}\xi^{a}_{L}$, $\kappa_{o} = \sum_{a=4}^{5} \xi^{a}_{R}\xi^{a}_{L}$, $\kappa_{I} =  \xi^{6}_{R}\xi^{6}_{L}$, and $\kappa_{c} = \sum_{a=7}^{8} \xi^{a}_{R}\xi^{a}_{L}$. The indices ``$s$'', ``$o$'', ``$I$'', and ``$c$''  refer to the SU(2)$_{s}$, U(1)$_{o}$, $\mathcal{Z}_{2}$, and U(1)$_{c}$ symmetries respectively.

Away from half filling, case (b), the charge mode is decoupled from the other modes, and we don't need to consider $g_{i= 6\sim9}$. Thus, we have 
\begin{equation}
\begin{split}
\mathcal{H} &= -i\frac{v}{2\pi} \sum_{a=1}^{6}\left( \xi^{a}_{R}\partial\xi^{a}_{R}- \xi^{a}_{L}\partial\xi^{a}_{L}   \right) \\
&+ \frac{g_{1}}{2} \kappa_{s}^{2} + g_{2} \kappa_{s}\kappa_{o} +g_{3}\kappa_{s}\kappa_{I} \\
&+ g_{4}\kappa_{o}\kappa_{I} +\frac{g_{5}}{2} \kappa_{o}^{2}.
\end{split}
\label{so6}
\end{equation}
As we mentioned before, even without an explicit U(1) orbital symmetry for the lattice Hamiltonian, the low energy theory has the effective $\widetilde{\text{U}(1)}_{{o}}$ symmetry.\cite{Lin1998,Controzzi2005,Nonne2010} Therefore the structure of the refermionized forms are the same as above. Thus, cases (c) and (d) now have the same form as Eqs. \eqref{so8} and \eqref{so6} with different values of $g$'s.\footnote{To get the same form, we have to redefine $\psi^{c\pi}_{L} = \frac{1}{\sqrt{2}} \left( \xi^{5}_{L} - i \xi^{6}_{L}\right)$. Here the modes correspond to the new orbital part are $\xi_{5}$ and $\xi_{6}$, and the Ising mode is carried by $\xi_{4}$.} 

For the refermionized forms, duality mappings are defined as, $\xi^{a}_{L} \rightarrow - \xi^{a}_{L}$ while keeping right-moving parts untouched. It is easy to see that Eqs. \eqref{so8} and \eqref{so6} are invariant under such transformations if we change the signs of some coupling constants as well. To retain the form of the Hamiltonian, only mappings that transform all the Majorana fields in the same symmetry sector are permitted. For example, for the SU(2) spin part, we should map the three left Majorana fermions, $\xi^{a=1\sim 3}_{L}$, at the same time. For notational convenience, we define $\Omega_{\nu}$ as an operation of  $\xi^{a}_{L} \rightarrow - \xi^{a}_{L}$ for all the $a$'s in a symmetry sector $\nu$. With this at hand, it is obvious that allowed mappings for half filling cases are
\begin{equation}
\Omega_{O(8)} \equiv \{ \Omega_c, \Omega_o, \Omega_I,  \Omega_s,  \Omega_{c,o},  \Omega_{c,I},  \Omega_{o,I} \}.
\end{equation}
The number of independent mappings is 3, and other mappings just follow from them e.g. $\Omega_{o,I} = \Omega_{o}\Omega_{I}$. Away from half filling, the charge mode is separated, so only 3 of the above mappings are left,
\begin{equation}
\Omega_{O(6)} \equiv \{ \Omega_o, \Omega_I,  \Omega_s \},
\end{equation}
and two of them are independent. An immediate consequence of these dualities and DSE is that although we showed 16 insulating phases for half filling systems, and 8 metallic phases for incommensurate filling, only a part of them are realized.

Now, we will show such possible ground states for each model. We start from the ``fundamental" SO(8) GN model, 
\begin{equation}
\mathcal{H} = -i\frac{v}{2\pi} \left( \vec{\xi}_{R}\partial \vec{\xi}_{R}- \vec{\xi}_{L}\partial \vec{\xi}_{L}   \right) + \frac{g}{2} \left( \vec{\xi}_{R} \vec{\xi}_{L} \right)^{2},
\end{equation}
which appears at low-energy when all the $g$'s in Eq. \eqref{so8} converge to the same value as a result of DSE. For case (a), this model represents the BDW phase, and other possible phases are found by applying $\Omega_{O(8)}$ to the BDW state (see Table \ref{ground states}). We denote them as $\Gamma_{y}$, and they are 
\begin{multline}
\Gamma_{y}: \text{BDW, CDW, $p'$CDW, $p'$BDW,}\\
\text{S-Mott, S'-Mott, RT, HO}.
\label{new}
\end{multline}
The case (b) follows from the relation between insulating states and metallic states (Table \ref{doping}), or applying  $\Omega_{O(6)}$ to the CDW phase, which is ``fundamental". The original lattice model we are considering here is invariant under the orbital U(1) rotation about $y$-axis. As we mentioned in Sec. \ref{bosonization}, the generator of this symmetry cannot be expressed by a single local bosonic field within Abelian-bosonization scheme.

This combination, $\Gamma_{y}$, is different from the ones which have been studied extensively; previously studied phases are
\begin{equation}
\Gamma_{z}: \text{BDW, CDW, $p$BDW, $p$CDW, RS, HC, RT, HO}
\label{81}
\end{equation}
and
\begin{multline}
\widetilde{\Gamma_{z}}: \text{S-Mott, S'-Mott, D-Mott, D'-Mott,} \\
\text{$s'$CDW, $s'$BDW, $p'$BDW, $p'$CDW}.
\label{82}
\end{multline}
The former, $\Gamma_{z}$, appears in models with weak transverse hopping, and with the U(1)$_{o}$ symmetry about $z$-axis.\cite{Lee2004, Nonne2010} The latter, $\widetilde{\Gamma_{z}}$, appears when the model has strong transverse hopping, and the low energy theory possesses the $\widetilde{\text{U(1)}_{o}}$ symmetry;\cite{Balents1996, Lin1998,Tsuchiizu2002, Wu2003,Chudzinski2008} our cases (c) and (d) belong to this category (see Table \ref{ground states}). 

\begin{figure}[!tb]
\centering
\includegraphics[scale=1]{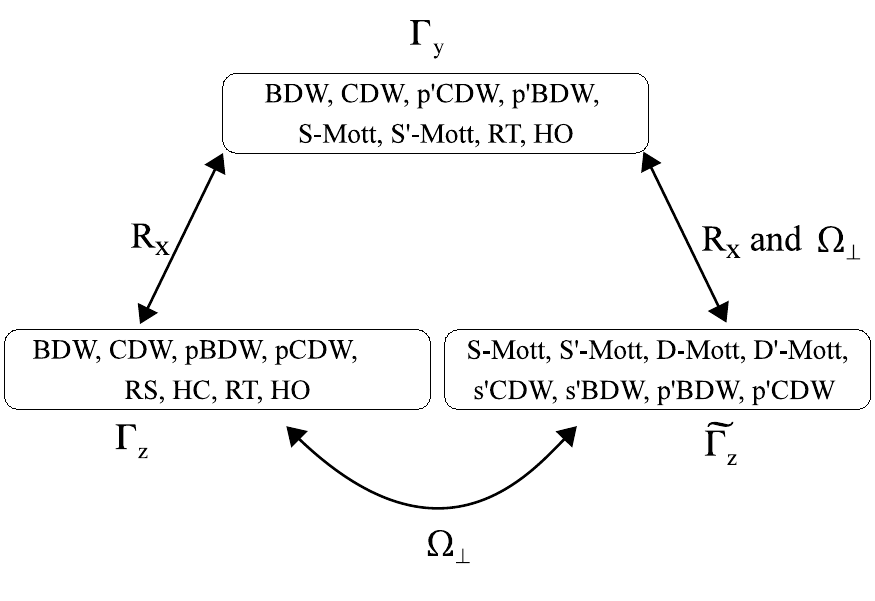}
\caption{Relationships among three groups of insulating ground states. $R_{x}$ indicates the rotation about $x$-axis in orbital space in Eq. \eqref{rx}, and $\Omega_{\perp}$ is strong-weak tunneling duality mapping in Eq. \eqref{sw}. For example, the S'-Mott state in $\Gamma_{y}$ maps to the HC state in $\Gamma_{z}$ by $R_{x}$.}
\label{three}
\end{figure}

The connection between $\Gamma_{y}$ and $\Gamma_{z}$ is obvious. Since the generator of the orbital symmetry for each case is $y$- or $z$-component of Eq. \eqref{orbital}, they are simply mapped to each other by a rotation around $x$-axis:
\begin{equation}
R_{x}:
\begin{pmatrix}
 c_{rAs}'\\ 
 c_{rBs}'
\end{pmatrix} 
= \frac{1}{\sqrt{2}}
\begin{pmatrix}
1 & -i \\ 
-i& 1  
\end{pmatrix} 
\begin{pmatrix}
 c_{rAs}\\ 
 c_{rBs}
\end{pmatrix} .
\label{rx}
\end{equation}
This transformation does not affect the charge and spin generators. For instance, the S'-Mott state in $\Gamma_{y}$ goes to the HC state in $\Gamma_{z}$ by $R_{x}$. The correspondence among other states is given in Fig. \ref{three}. On the other hand, $\widetilde{\Gamma_{z}}$ and $\Gamma_{z}$ transform each other by so-called \textit{strong-weak tunneling duality},\cite{Controzzi2005, Nonne2010}
\begin{equation}
\Omega_{\perp}: \ \ c_{Lm\uparrow} \rightarrow c^{\dagger}_{Lm\downarrow}, \ \  c_{Lm\downarrow} \rightarrow - c^{\dagger}_{Lm\uparrow}.
\label{sw}
\end{equation}
The relationships among metallic ground states are easily obtained by Table \ref{doping}.

Therefore, we found that in addition to underlying band structure, the form of interaction also affects the possible combinations of ground states. We summarized these results in Table \ref{ground states} and Fig. \ref{three}.

\subsection{Quantum phase transition}
The quantum phase transitions among gapped ground states could be either first order or second order. For the transitions among states connected by duality, the modes which are not involved in the mapping become massive at higher energy, and the effective low energy theory near the transition contains only Majorana fields flipped by the mapping.\cite{Boulat2009}

For a single Majorana field, it becomes the critical Ising model,
\begin{equation}
\mathcal{H} = -i\frac{v}{2\pi} \left( \xi_{R}\partial\xi_{R}- \xi_{L}\partial\xi_{L}   \right)-im \xi_{R}\xi_{L}
\end{equation}
Over the transition, the mass changes its sign, and this represents a second-order phase transition. With more than one field, the low-energy effective theory becomes the massive O(N) GN model,
\begin{equation}
\mathcal{H} = -i\frac{v}{2\pi} \left( \vec{\xi}_{R}\partial \vec{\xi}_{R}- \vec{\xi}_{L}\partial \vec{\xi}_{L}   \right)-im \vec{\xi}_{R} \vec{\xi}_{L} + \frac{g}{2} \left( \vec{\xi}_{R} \vec{\xi}_{L} \right)^{2}.
\label{OGN}
\end{equation}
The fate of further renormalization to lower energy determines whether the phase transition is first-order or second-order depending on the final fixed point for the critical fields.\cite{Gross1974,Shankar1985,Lin1998,Tsuchiizu2002,Controzzi2005} The transition line is defined as the point where the $m$ in Eq. \ref{OGN} goes to zero, and the critical fields are expressed by a massless GN model in the vicinity of transition. For $N=2$, it is known that the system can be mapped to a Gaussian model, so it is a second order transition. For $N \geq 3$, however, if the coupling constant in the GN model is positive ($g>0$), the renormalization flow departs to a strong coupling fixed point (asymptotic free), since RG equation is given by
\begin{equation}
\dot{g} \propto g^{2}.
\end{equation}
At this fixed point, the mass is generated dynamically, and the system is off-critical. We can see this either by mean-field treatment of the interaction (reducing the quartic part to quadratic with the order parameter, $\langle \xi_{R}\xi_{L}\rangle$), or by stationary phase approximation, which becomes exact when $N\rightarrow \infty$. At this massive fixed point, there are two degenerate minima about two signs of mass, and they correspond to two phases connected by this first-order transition. On the other hand, when $g < 0$, further renormalization reduces $g$ to 0, and the system reaches a massless fixed point; this represents a second-order transition.

When the transition is second order, the critical theory is described by a conformal field theory (CFT) due to its dimensionality, (1+1). Each CFT is characterized by its central charge, $c$, which roughly expresses the number of critical fields. $c=1/2$ is the $\mathcal{Z}_{2}$ Ising critical theory, $c=1$ is the U(1) Gaussian theory, and $c=3/2$ is the SU(2)$_2$ Wess-Zumino-Novikov-Witten theory. With the duality mappings, it is easy to read off the central charge of each CFT.  Since each Majorana fermion carries $c=1/2$, the number of fields flipped by a mapping directly tells us the central charge. We will identify the phase transitions appearing in our phase diagrams more precisely in the next section.

\section{RG equations and phase diagrams}
\label{rg}
In the RG equations obtained from Abelian bosonization, we use normalized coupling constants defined as
\begin{equation}
y_i \equiv \frac{g_i}{4\pi v}.
\end{equation}
RG equations are derived using the operator product expansion (OPE) \cite{cardy1996scaling, Balents1996, VonDelft1998} and integrating out higher frequency modes. The RG equations are complicated, so we will not show them explicitly. In all of the cases which we have examined, the RG equations may be expressed as
\begin{equation}
\frac{d y_{i}}{d l} = \frac{\partial V}{\partial y_{i}}.
\end{equation}
with a potential function, $V[y_{1}(l), y_{2}(l), \cdots, y_{n}(l)]$. The RG flow is to the valleys of the potential in the beginning and then along the valley. The potential structure is consistent with the arguments of Ref. \onlinecite{Chang2005}, which suggests that the presence of a potential function is related to Zamolodchikov's c-theorem.\cite{Zamolodchikov1986a} For the commensurate case, $k_{A}=k_{B} =\pi/2$, the potential is
\begin{widetext}
\begin{equation}
\begin{split}
V[y_{i}]&=  y_{1a}y_{1c}y_{\parallel c} - y_{1a}y_{1b}y_{2a} + y_{1a}y_{3a}y_{\parallel b}- y_{1c}y_{1d}y_{2c} + y_{1c}y_{3c}y_{\parallel b}-  y_{1d}y_{\parallel a} y_{1b}\\
&- y_{1d}y_{3b}y_{\parallel b}+ y_{2c}y_{3b}y_{3c} + y_{\parallel c}y_{3a}y_{3c}- y_{1b} y_{3d}y_{\parallel b}+ y_{2a}y_{3a}y_{3d}- y_{\parallel a}y_{3b}y_{3d}\\
&+ \frac{1}{2}\left( y_{3a}^{2}+ y_{3b}^{2}+y_{3c}^{2}+y_{3d}^{2}+y_{\parallel b}^{2} \right) y_{c0}
+ \frac{1}{2}\left( -y_{1c}^{2}- y_{2c}^{2}-y_{\parallel c}^{2}+y_{\parallel a}^{2}+y_{1b}^{2} +y_{2a}^{2} -y_{3c}^{2} + y_{3d}^{2}\right) y_{c\pi}\\
&+ \frac{1}{2}\left( y_{1d}^{2}+ y_{1a}^{2}+y_{1c}^{2}+y_{1b}^{2}+y_{\parallel b}^{2} \right) y_{s0}
+ \frac{1}{2}\left( y_{1d}^{2}- y_{1a}^{2}+y_{2c}^{2}-y_{\parallel c }^{2}+y_{\parallel a}^{2}- y_{2a}^{2} -y_{3a}^{2}  +y_{3b}^{2}\right) y_{s\pi},
\end{split}
\label{potential}
\end{equation}
\end{widetext}
where we introduced $K_{\mu \nu} \equiv 1 -y_{\mu \nu}$ for each Luttinger parameter. For $k_{A}\neq k_{B}$ or doped cases, we should remove some coupling constants which are not allowed by momentum conservation. The RG equations valid even when velocities are different are given in Appendix \ref{bf}. We checked that both RG equations give consistent results when $v_{A} =v_{B}$. 

To get phase diagrams, we integrated the RG equations numerically until one of the coupling constant becomes of the order of $1 (\equiv t)$. We used initial values of coupling constants as small as $10^{-8}$--$10^{-3}$. Due to the hidden potential structure, the asymptotic behavior of the RG flow is captured by the following ansatz,\cite{Balents1996,Chang2005}
\begin{equation}
g'_{i}[l] = \frac{g_{0i}}{l-l_c},
\label{ray}
\end{equation}
where $l_c$ is the length at which the relevant couplings diverge, and the $g_{0i}$ determines the ratio among them. This represents the fixed ray of relevant coupling constants. Then, the bosonic fields are pinned down to the minima of the effective potential. These values enable us to determine the order parameter which takes a non-zero value.

The obtained phase diagrams are shown in Figs. \ref{a}--\ref{e}. They correspond to the various cases in Table \ref{band}. We also investigated the effect of velocity anisotropy to the phases in physically relevant parameter region, $0<J<U/2$,  using the RG equations based on the fermionic Hamiltonian. We studied the range, $1 \leq v_{A}/v_{B} \leq 10$. 

At various points in the following discussion, we also label the phases in the ``$CnSm$'' notation, which indicates $n$ massless charge modes and $m$ massless spin modes introduced in Ref. \onlinecite{Balents1996}. 

\subsection{$k_{A} = k_{B} = \pi/2$ at half filling}
\begin{figure}[!htb]
\centering
\subfigure[$v_{A}/v_{B} =1$]{
\includegraphics[scale=0.48]{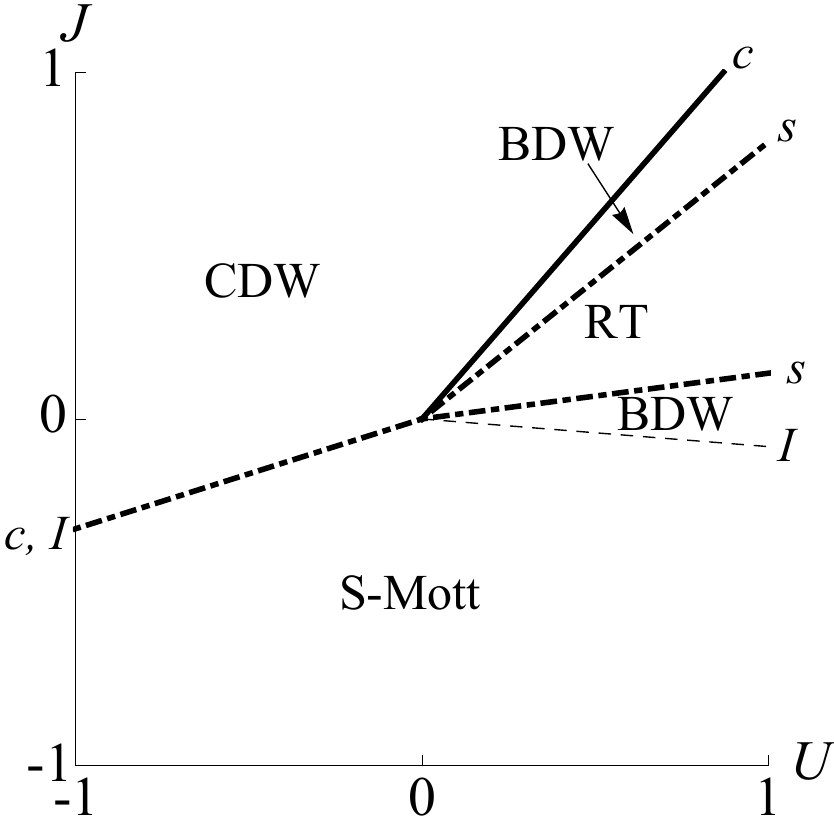}}
\subfigure[$v_{A}/v_{B} =5$]{
\includegraphics[scale=0.48]{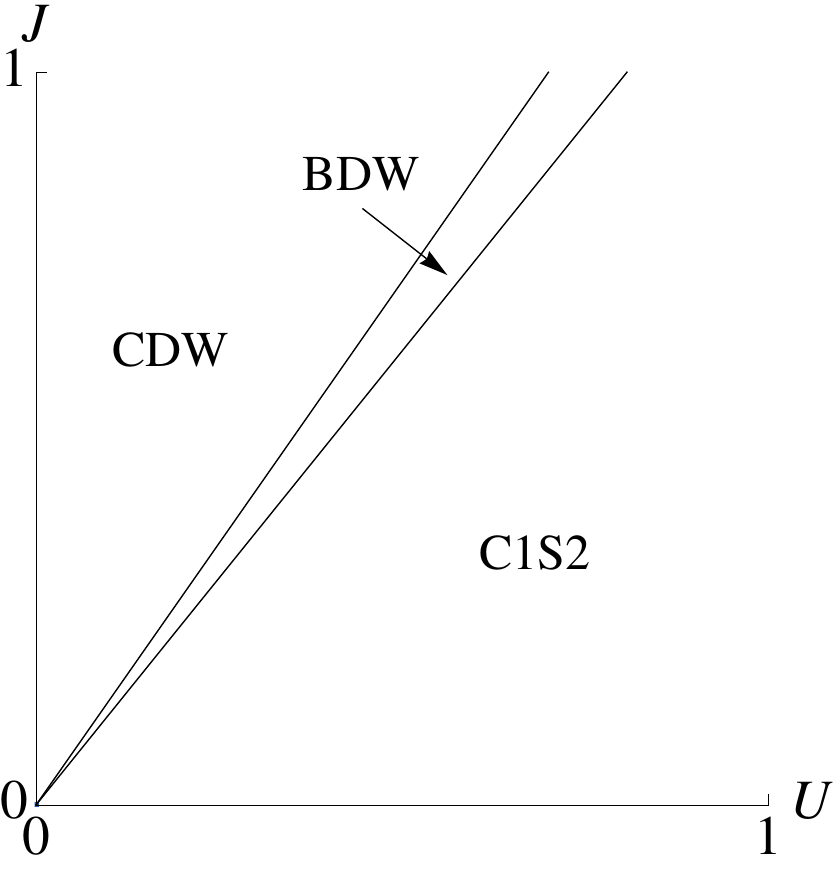}
\label{av}}
\caption{Phase diagram for case (a), $k_{A} = k_{B} =\pi/2$ with equal velocities. Physically relevant region is $0<J<U/2$. A dashed line indicates SU(2) or first order transition. A solid line is a Gaussian theory, and a dotted line is an Ising transition. ``$c$'', ``$o$'', ``$s$'', and ``$I$'' indicate the critical fields on the transition line. All of these phases are $C0S0$. }
\label{a}
\end{figure}
First, we look at case (a), where $k_{A}=k_{B}$ at half filling (Fig. \ref{a}). Absence of the HO phase indicates that the system does not flow to the enlarged orbital SU(2) symmetric state, although this could happen in principle by changing initial conditions. The RT phase, a high spin sate, resides in $0<J< U$ region, which may be accessible in real material. 

Although precise boundaries do not coincide exactly, the corresponding Hartree-Fock (HF) phase diagram shows similar structure.\cite{Okamoto2011} There we have the SDW phase instead of the RT state in $0<J<U$; they are both locally high-spin configurations with anti-ferromagnetic orders along the chain. The S-Mott state found in the bosonization result corresponds to mainly the degenerate state of the $s'$CDW and $p$CDW orders with a smaller region of $s$SS in the HF phase diagram. In the bosonic language, the order parameter of the $s$SS phase is the same as that of the S-Mott state except the total charge mode. The $s'$CDW and $p$CDW states are degenerate due to the orbital symmetry, and they are Ising dual to the S-Mott and Haldane orbital (HO) phases respectively (see Table \ref{doping}). The HO phase does not appear in the bosonic calculation, but it is connected to the S-Mott phase by the duality mapping of the orbital sector (see Table \ref{ground states}). The CDW state stays almost the same regime in both phase diagrams. The BDW phase does not appear in the HF phase diagrams, since its HF energy is higher than that of the CDW phase.

The critical properties of the CDW-BDW transition are given by a U(1) Gaussian theory of the charge sector, while those of the BDW-S-Mott transition are given by the $\mathcal{Z}_{2}$ Ising theory. The rest of transitions are either SU(2) criticality or first order; the critical fields at the BDW-RT transition line are spin modes, and at the CDW-S-Mott line are the charge and Ising fields. 

The phase transition from the RT state to the CDW state with increasing $J$ is similar to the SDW-CDW transition found in the extended Hubbard model (EHM).\cite{Nakamura2000,Tsuchiizu2002a} The EHM has a nearest neighbor interaction, $V n_{j} n_{j+1}$, in addition to the Hubbard interaction, $U n_{j \uparrow}n_{j, \downarrow}$. As the former interaction becomes predominant, particles try to form a CDW state, while strong $U$ prefers SDW. In the weak coupling regime, it is found that the SDW state undergoes a spin-gap transition to a BDW state, and then becomes the CDW state through a Gaussian transition of the charge sector. In the strong-coupling regime, these two transition lines are coupled to a first order transition line. In our model, strong $J$ plays the same role as $V$ in the EHM; large $J$ induces an attractive on-site interaction [see Eq. \eqref{int}] leading to the CDW state. The properties of transitions from the RT phase to the CDW phase, and the existence of the narrow BDW region are also the same as in the EHM. Therefore, we expect that in the strong-coupling regime, the RT-CDW transition in our model also becomes first order, although this has not been demonstrated.

Now, we consider the effect of velocity difference in the first quadrant; $U, J >0$ [Fig. \ref{av}]. As $v_A / v_B$ becomes as large as 1.5, we found that the RT and BDW states in small $J>0$ are completely replaced by a $C1S2$ state, where only a charge mode of a single band becomes massive, and the rest is massless. The CDW and BDW states in $J >U>0$ are robust to the change in velocity. This is because the large anisotropic velocities suppress the interband scattering, resulting in the domination of intraband scattering. As $v_A / v_B$ is increased beyond 1.5, the $C1S2$ phase becomes larger, although the BDW phase always exists between the CDW and $C1S2$ phases.

\subsection{$k_{A} = k_{B}$ at incommensurate filling}
\begin{figure}[!htb]
\centering
\subfigure[$v_{A}/v_{B} =1$]{
\includegraphics[scale=0.48]{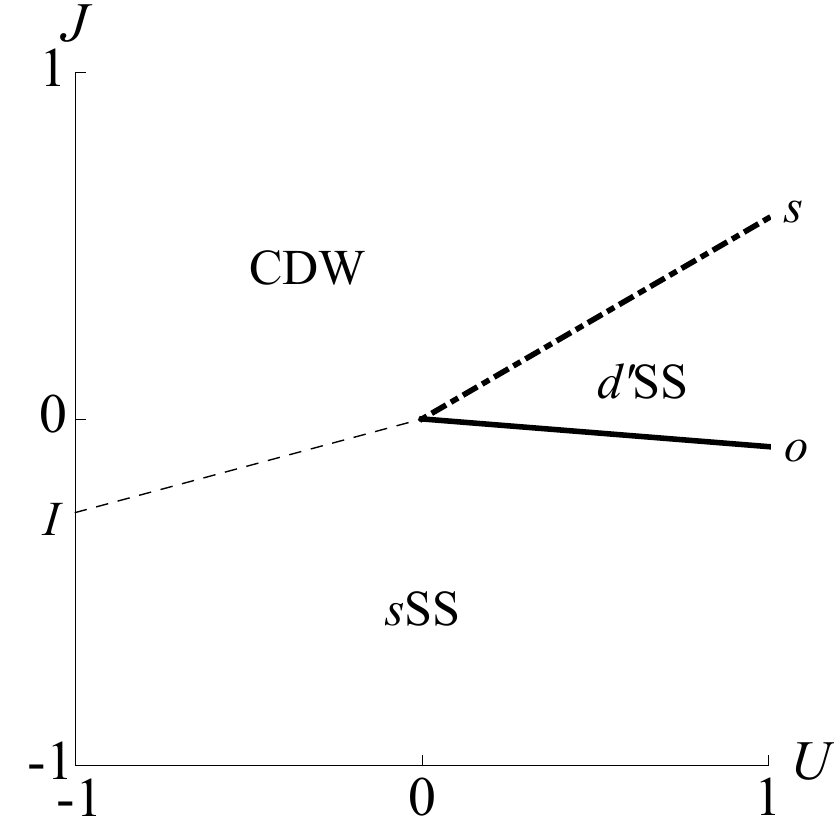}}
\subfigure[$v_{A}/v_{B} =5$]{
\includegraphics[scale=0.48]{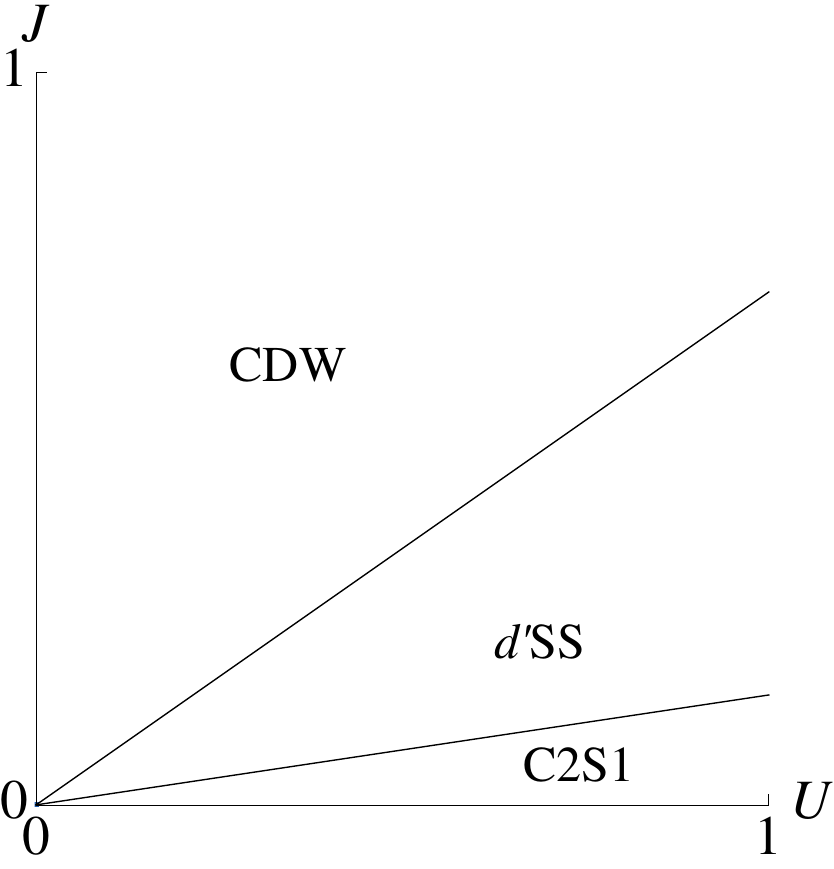}
\label{bv}}
\caption{Phase diagram for case (b), $k_{A} = k_{B}$ away from half filling with equal velocities. Physically relevant region is $0<J<U/2$. A dashed line indicates SU(2) or first order transition. A solid line is a U(1) Gaussian theory, and a dotted line is an Ising transition. ``$c$'', ``$o$'', ``$s$'', and ``$I$'' indicate the critical fields on the transition line. All of these phases are $C1S0$.}
\label{b}
\end{figure}
The phase diagram for $k_{A}=k_{B}$ with incommensurate filling is given in Fig. \ref{b}. This is similar to the one at half filling if we replace the insulating states to corresponding metallic ones; S-Mott to $s$SS, RT to $d'$SS, and BDW to CDW. The BDW sate near $J\simeq 0$ in Fig. \ref{a} disappears. The transition between the CDW and $d'$SS states is governed by spin modes leading to the SU(2) criticality or first order transition. The CDW-$s$SS transition is an Ising transition, $c=1/2$. Finally the $d'$SS-$s$SS transition is described by a Gaussian theory of the orbital sector.

The velocity difference in a quadrant, $U, J > 0$, does not modify the large $J$ regime, though a $C2S1$ state appears at small $J$ [Fig. \ref{bv}]. The $C2S1$ phase was observed in other two-leg ladder systems when the velocity difference becomes large.\cite{Balents1996, Chudzinski2008}

The HF phase diagram of this case\cite{Okamoto2011} has the SDW phase for $0<J<0.6U$, which corresponds to the $d'$SS phase found by bosonization; both of them are locally high-spin states. In the negative $J$ region, we have the $s$SS state in Fig. \ref{b}, while the HF calculation gives not only the $s$SS state, but also a large region of the $s'$CDW phase. As we mentioned, this CDW state is Ising dual to the S-Mott phase, which is the insulating analogue of the $s$SS state. In the large $J>0$ region, we found $p_{y}'$-wave spin-triplet superconductivity in the HF phase diagram, which is replaced by CDW in the bosonic calculation. The CDW state around $0<J< -U$ is robust, and we observe it both at HF level and after renormalization.

\subsection{$k_{A} \neq k_{B}$ at half filling}
\begin{figure}[!htb]
\centering
\includegraphics[scale=0.48]{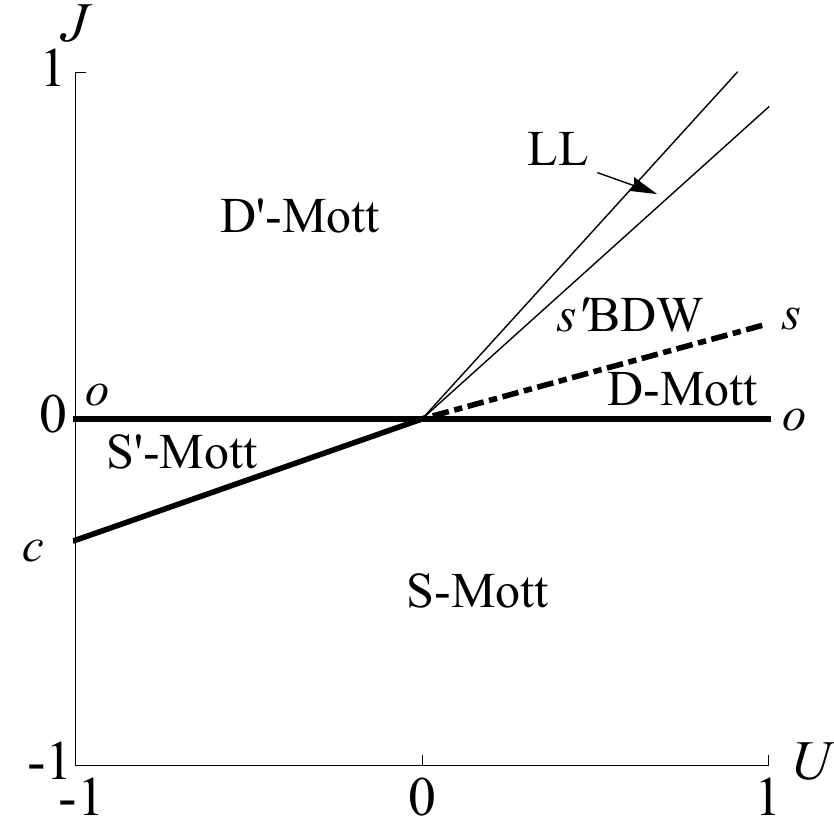}
\caption{Phase diagram for case (c), $k_{A} \neq k_{B}$ at half filling with equal velocities. Physically relevant region is $0<J<U/2$. A dashed line indicates SU(2) or first order transition. A solid line is a U(1) Gaussian theory. A thin line indicates KT transition. ``$c$'', ``$o$'', ``$s$'', and ``$I$'' indicate the critical fields on the transition line.}
\label{c}
\end{figure}
For the system at half filling, but with two different Fermi momenta, the phase diagram is given in Fig. \ref{c}. There is a narrow Luttinger Liquid (LL) phase near $U\simeq J>0$, where all the modes are massless. The transition between massive phases and a LL phase is Kosterlitz-Thouless (KT) in the sense that a LL phase is critical and has power-low correlation, while massive phases have exponentially decaying correlations. The transitions among Mott phases all have a Gaussian criticality; the S-Mott-S'-Mott transition is governed by the charge sector, and others are by the orbital sector. On the other hand, the $s'$BDW-D-Mott transition line is the SU(2) criticality or first order. In this case, the velocity difference does not modify the phase digram in physically relevant region essentially.

The corresponding HF phase diagram shows the SDW and $s'$SDW states in the physically relevant region,\cite{Okamoto2011} while they are replaced by the $s'$BDW and D-Mott phases in the RG phase diagram. This is a notable difference between the result of two equivalent bands and that of inequivalent bands. For the $k_{A} = k_{B}$ cases, locally high-spin states, RT and $d'$SS, are dominant in $0<J<U/2$, while low-spin configurations, $s'$BDW and D-Mott, are found in the $k_{A} \neq k_{B}$ case. We understand these low-spin states as a result of decoherence by two different wave numbers. In essence, density waves with different phases in the two bands mean that the energy contribution from the $J$ interaction averages out to zero. The CDW phase, which dominates large positive $J$ region at mean-field level, is replaced by the D'-Mott state after renormalization. For this case, the $p$SS state, the metallic analogue of the D'-Mott state, is subdominant with large positive $J$ at the HF level, and  this is more enhanced than the CDW order during the renormalization flow. Negative $J$ region of the HF phase diagram is again covered by the $s$SS and $s'$CDW phases, which are related to the S(')-Mott state.

\subsection{$k_{A} \neq k_{B}$ at incommensurate filling}
\begin{figure}[!htb]
\centering
\includegraphics[scale=0.48]{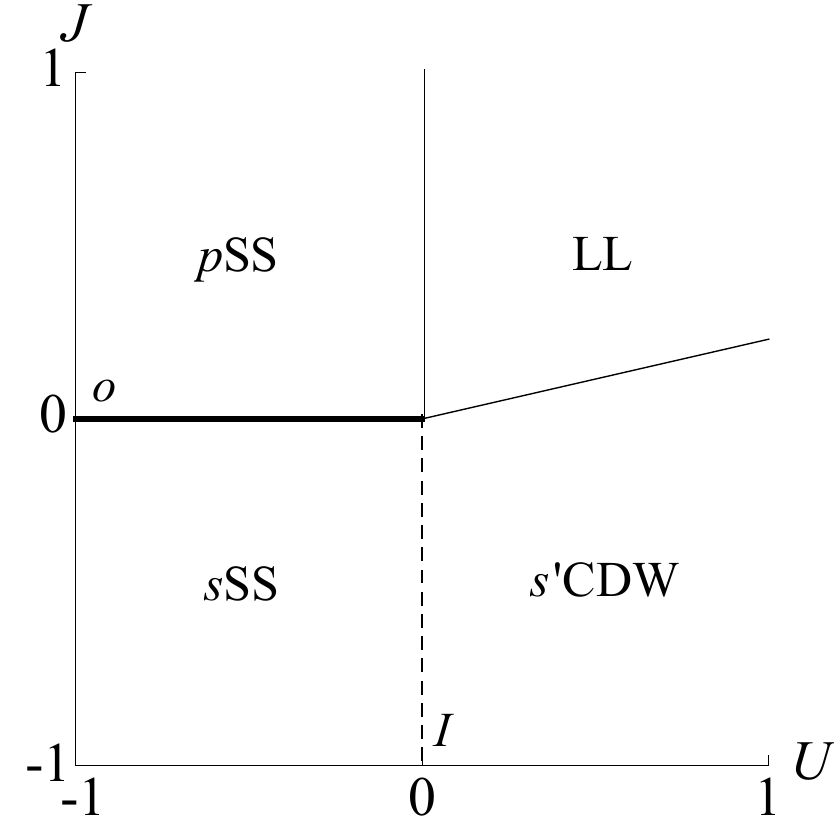}
\caption{Phase diagram for case (d), $k_{A} \neq k_{B}$ away half filling with equal velocities. Physically relevant region is $0<J<U/2$. A solid line is a U(1) Gaussian theory, and a dotted line is an Ising transition. A thin line is KT transition. ``$c$'', ``$o$'', ``$s$'', and ``$I$'' indicate the critical fields on the transition line.}
\label{d}
\end{figure}
With inequivalent Fermi momenta and incommensurate filling (Fig. \ref{d}), the physically relevant region is covered by a LL phase,  and the $s'$CDW phase. The $p$SS and $s$SS phases can be understood as the reminiscent of the D'-Mott and S(')-Mott states which exist at half filling. Again, transitions between the LL phase and massive phases are KT type except the total charge mode remaining massless in both phases. The phase transition between the $p$SS and $s$SS states is governed by a U(1) Gaussian criticality of the orbital sector. The transition between the $s'$CDW and $s$SS phases is Ising type. As the velocity anisotropy becomes larger, the $s'$CDW phase is gradually suppressed, and whole area in physically relevant region is covered by the LL state for $v_A /v_B \geq6$.

The HF phase diagram in this case is similar to the RG phase diagram. We have the $p(s)$SS state with large negative $U$ and small positive (negative) $J$. For large negative $J$, we have the $s'$CDW state. The $p$TS and SDW states appearing in positive $J$ regime of the HF phase diagram are renormalized to the Luttinger liquid phase.

\subsection{$k_{A} =\pi/2 \neq k_{B}$ at incommensurate filling}
\begin{figure}[!htb]
\centering
\includegraphics[scale=0.48]{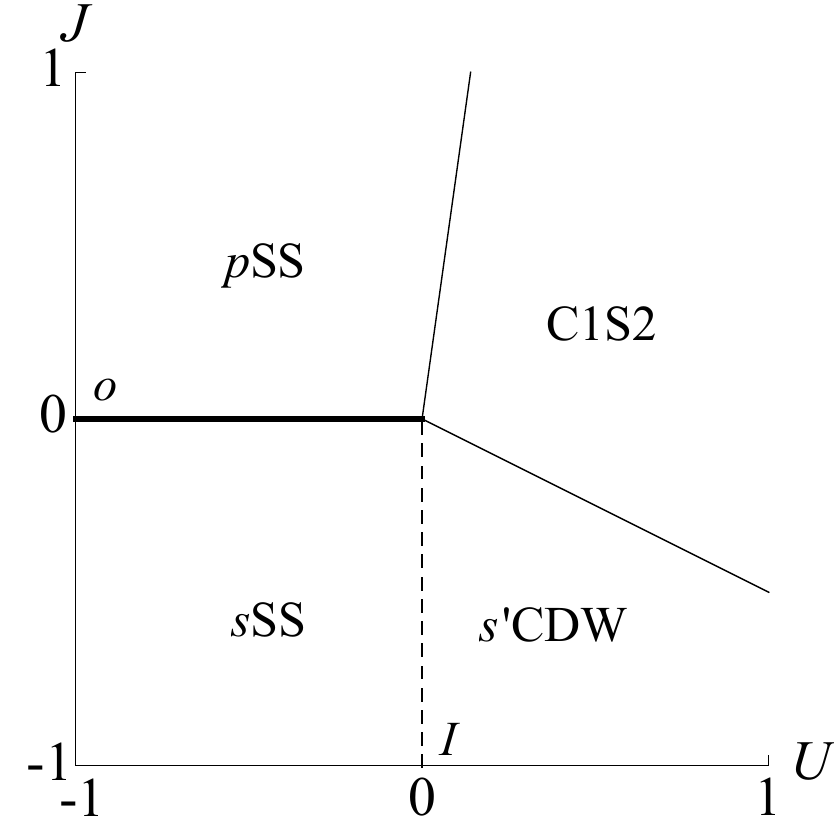}
\caption{Phase diagram for case (e), $k_{A} =\pi/2 \neq k_{B}$ away from half filling with equal velocities. Physically relevant region is $0<J<U/2$. A solid line is a U(1) Gaussian theory, and a dotted line is an Ising transition. A thin line is KT transition. ``$c$'', ``$o$'', ``$s$'', and ``$I$'' indicate the critical fields on the transition line.}
\label{e}
\end{figure}
The phase diagram of an orbital selective Mott case (Fig. \ref{e}) is almost similar to that of case (d) except Luttinger liquid state is replaced by a $C1S2$ phase, where a commensurate band opens a charge gap, and the rest of the modes remain massless. The velocity anisotropy in both directions, $v_A / v_B <1$ and $v_A / v_B >1$, does not modify the $C1S2$ state in the region, $0<J<U/2$. The HF phase diagram for this case is almost the same as the previous case with $k_{A} \neq k_{B}$ at incommensurate filling.

\section{Fully spin-polarized case: Charge-Orbital model}
\label{strong}
So far, we have limited our focus to the weak coupling limit. However, in multiorbital systems the strong-coupling limit can bring qualitatively new effects. \cite{Kugel1972, Cyrot1975, Gill1987, Sakamoto2002, Okamoto2011} In particular, in the physical ($J>0$) case, one may expect every ion to be in the state of maximal spin consistent with given total occupation. In this circumstance, we expect antiferromagnetic order in the half-filled case, and when the system is slightly doped, more complicated structures such as phase separation and spiral phases will appear. At filling further away from $n=2$, the system shows ferromagnetism (FM) with an orbital order. Considering the fact that ferromagnetic states dominate the phase diagram at most fillings,\cite{Sakamoto2002, Okamoto2011} in this section, we investigate the possible orbital orders assuming that the system is fully spin polarized. In other words, we consider the effect of residual backscattering in the subspace of the charge and orbital sectors assuming the spin excitations are frozen. We leave the investigation of the regime close to half filling for future study.

Suppose that all the electrons have the same spins, the model in Eq. \eqref{kin} and Eq. \eqref{int} is then reduced to,
\begin{equation}
H = H_\text{kinetic}+\sum_{i}  \left( U-3J \right)n_{i A }n_{i B},
\label{co}
\end{equation}
where we omit the spin index. Now the SU(2)$_{s}$ symmetry is lost, but we can regard the orbital part as pseudo spins. If the two bands have the same $k_{F}$ and velocities, the system has the orbital SU(2) symmetry. The band splitting in orbital sector is isomorphic to the Zeeman splitting by magnetic field. Therefore, the model now turns to a simple Hubbard model with effective interaction $U_\text{eff}= U-3J$, with or without magnetic field. 

Depending on the effective crystal field splitting between the two orbitals $\Delta$, and the band widths, there may be three different scenarios in this model. The first case is that the two bands are completely degenerate: $\Delta =0$, and $k_{A} =k_{B}$. We expect a staggered orbital order to appear. On the other hand, when there exists either small splitting or when the band widths are slightly different, the two momenta are not equal, and orbital orders might be suppressed. We will discuss these scenarios below using bosonization method. However, there is another scenario, which may arise when either splitting is large, or two band widths are greatly different. Then, only one band has states at the Fermi surface and the physics is trivial.

\begin{figure}[!tb]
\centering
\subfigure[$k_{A} = k_{B}$]{
\includegraphics[scale=0.55]{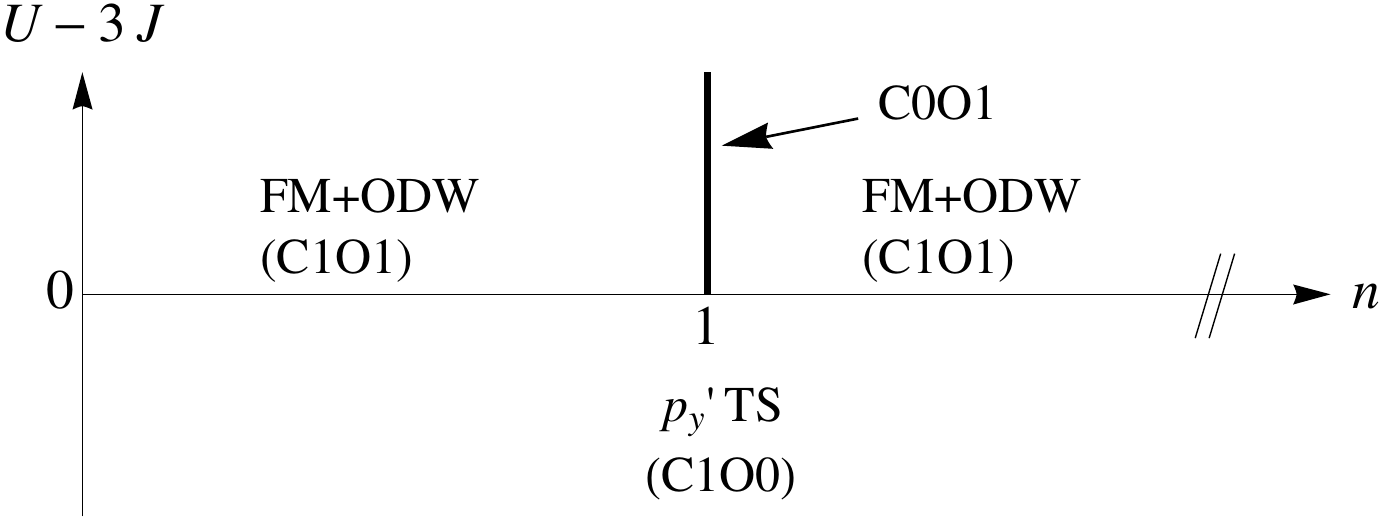}
\label{fm1}}
\subfigure[$k_{A} \neq k_{B}$]{
\includegraphics[scale=0.55]{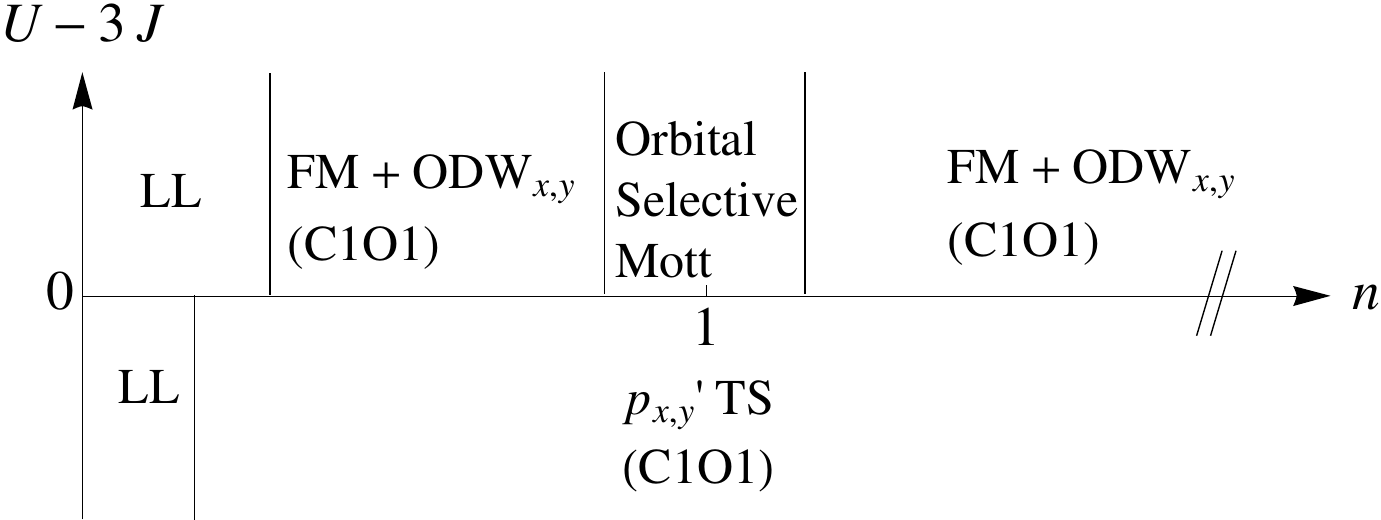}
\label{fm2}}
\caption{Schematic phase diagrams for the spin-polarized model for two equivalent bands or slightly different bands. When $k_{A} \neq k_{B}$, the ODW$_{x}$ and ODW$_{y}$ phases have the same correlation exponent, and so do $p_{x}'$TS and $p_{y}'$TS. Filling closer to $n=2$ is not investigated. ``$CnOm$ '' represents a state with $n$ massless charge modes and $m$ massless orbital modes.}
\label{fm}
\end{figure}

We first look at the degenerate case, $k_A = k_B$, corresponding to the absence of magnetic field. The bosonized form of the Hamiltonian in Eq. \eqref{co} at half filling is given by
\begin{equation}
\begin{split}
\mathcal{H} &=\frac{1}{2\pi}\sum_{\substack{\nu = c,o}} v_{\nu} \left[ K_{ \nu} ( \nabla \theta_{\nu} )^2 + \frac{1}{K_{\nu} }( \nabla \phi_{\nu} )^2 \right]\\
&+\frac{2}{(2\pi \alpha)^{2}} U_\text{eff} \cos \left(2\sqrt{2}\phi_{o}\right) \\
&-\frac{2}{(2\pi \alpha)^{2}} U_\text{eff} \cos \left(2\sqrt{2}\phi_{c}\right),
\end{split}
\label{fm_boson}
\end{equation}
where the Luttinger parameters and velocities are,
\begin{gather}
K_{c}v_{c} = K_{o}v_{o}=v\\
v_{c(o)}/K_{c(o)}=v\left( 1\pm \frac{U_\text{eff}}{\pi v} \right).
\end{gather}
Thus, the charge and orbital modes are decoupled, and each mode has a SU(2) symmetry. The total symmetry is $\text{SU(2)}\times\text{SU(2)}=\text{SO(4)}$.

Translating the analysis for the Hubbard model\cite{giamarchi2004quantum} to our charge-orbital model, we found the following results. For incommensurate filling, the last Umklapp term in Eq. \eqref{fm_boson} vanishes, and the charge mode is massless. Also the SU(2)$_{c}$ symmetry is broken to U(1)$_{c}$. About the orbital sector, we find:
\begin{description}
\item[(1) $U_\text{eff}>0$] Orbital density wave (ODW) has the longest correlation, and both orbital and charge modes are massless. The SU(2)$_{o}$ symmetry requires ODW about all three directions ($x, y, z$) are degenerate.
\item[(2) $U_\text{eff}<0$] The orbital sector becomes massive, and the phase with slowest decaying correlation is orbital-singlet superconductivity with parallel spins i.e. $p_{y}'$TS in Ref. \onlinecite{Okamoto2011}.
\end{description}
At half filling, the charge mode becomes massive ($K_{c} =1/2$) when the effective interaction is repulsive; the system is Mott insulating. The orbital part still gives ODW, and this FM+ODW state in $U>3J$ regime is observed both analytically\cite{Kugel1972, Cyrot1975, Okamoto2011} and numerically.\cite{Gill1987, Sakamoto2002} For the attractive side, the charge mode is gapless ($K_{c} =1$) with an orbital gap by the $p_{y}'$TS order; this is the Luther-Emery phase. This triplet superconductivity agrees with the numerical result by Sakamoto \textit{et al.}.\cite{Sakamoto2002} The results are summarized in Fig. \ref{fm1}.

Now we turn to the case with $k_{A} \neq k_{B}$; there is a pseudo magnetic field acting on the orbital space. At very small filling, only a single band is filled, so the ground state is ferromagnetic Luttinger liquid of a single gapless mode. When we dope enough the two bands start to share the Fermi surface. The SU(2)$_{o}$ symmetry is reduced to U(1)$_{o}$, and the $\cos \left(2\sqrt{2}\phi_{o}\right)$ term vanishes due to two different Fermi momenta. Thus, the orbital sector is always massless. With an attractive interaction, the band degeneracy occurs with smaller filling than with a repulsive interaction, since we guess the upper band is pulled down by the lower filled band for $U_\text{eff} < 0$. The charge mode is massive (massless) for repulsive (attractive) interaction at half filling. At tree-level, the states with longest correlations are ODW$_{x}$ and ODW$_{y}$ for $U_\text{eff} >0$, and $p_{x}'$TS and $p_{y}'$TS for $U_\text{eff} <0$. Since the orbital symmetry is explicitly broken, the exponents of correlation may differ for different directions. Finally, contrary to the complete degenerate case,  we speculate that an orbital selective Mott phase appears near $n=1$ for inequivalent bands; once one band is half filled, the commensurate wave vector opens a gap, and the other band remains metallic. Further filling just goes to the metallic band until it reaches half filling. Fig. \ref{fm2} presents the general phase diagram for this case.

\section{Conclusion}
\label{conclusion}
In this paper, we investigated a two-orbital Hubbard model which may encapsulate phenomena realized in transition-metal nanowires. Along with many aspects of two-leg ladder models, unique properties of transition-metal $d$-orbitals lead to several new results. 

In our analysis, we used the ideas of dynamical symmetry enlargement\cite{Lin1998,Konik2002,Controzzi2005} and duality relations\cite{Boulat2009,Nonne2010} to list 8 insulating phases at half filling, and 6 metallic phases away from half filling. Each phase is represented by a Gross-Neveu model at low energy, and phases are related to each other by duality mappings. The same analysis was done for weak-tunneling models with the U(1)$_{o}$ symmetry about $z$-axis,\cite{Nonne2010, Lee2004} and for strong-tunneling models with the $\widetilde{\text{U(1)}_{o}}$ symmetry. \cite{Lin1998,Tsuchiizu2002, Wu2003}. Some cases we studied in this paper belong to the latter category, and we obtained the same results. However, we also studied a model with a different orbital symmetry, the U(1)$_{o}$ symmetry about $y$-axis, and found that this form of  interaction gives rise to a new combination of ground states; half of which were found in weak-tunneling systems and the other half were in strong tunneling systems. This is because the underlying orbital rotational symmetry involves a generator which cannot be written by a single local bosonic field. The quantum phase transitions among those ground states were also briefly summarized using Gross-Neveu models as low-energy effective theories.\cite{Tsuchiizu2002, Boulat2009} The new combinations of ground states we found means that the model we proposed can exhibit new kinds of quantum phase transitions.

After classifying the possible ground states, we determined the ground state of the model by numerically integrating the RG equations. The phase diagrams are presented in Figs. \ref{a}--\ref{e}. The essential results for physically relevant regime, $0< J<U/2$, are as follows:
\begin{description}
\item[(1)] 
For two equivalent bands at half filling, we found a high spin state, RT, for relatively large $J$, and a low spin state, BDW, for small $J$ (Fig. \ref{a}). Keeping $k_{A} =k_{B}$ but introducing a velocity difference causes these phases to be replaced by a $C1S2$ state where only a charge mode in a single band becomes massive. Upon doping, the RT state turns into the $d'$SS state and the BDW state disappears (Fig. \ref{b}). A velocity difference leads to a $C2S1$ phase for small $J$, but the $d'$SS state still survives in large $J$.
\item[(2)]
For the cases with two different Fermi momenta, similar RG analyses have been done.\cite{Balents1996, Schulz1996,Tsuchiizu2002, Wu2003} Our main contribution for this case is that we identify the ground states for our specific model and interactions. The ground state in the physically relevant region is the $s'$BDW state for large $J$ and the D-Mott state for small $J$ at half filling (Fig. \ref{c}). These phases are robust to velocity differences. The surprising result here is that we have locally low-spin configurations even when $J>0$, while completely degenerate bands give high-spin states, RT and $d'$SS, in the same parameter regime. We think the low-spin configurations are achieved by the destructive interference between two different Fermi momenta, and therefore, the complete orbital degeneracy is crucial to have locally high-spin states. When the system is away from half filling, a Luttinger liquid phase becomes dominant, although we observed the $s'$CDW phase for very small $J$ (Fig. \ref{d}). This density wave state is wiped away by velocity difference, and replaced by the Luttinger liquid state. When only a single band is commensurate, the system exhibits an orbital-selective charge-gapped state due to the intraband Umklapp process while all the other modes remain massless (Fig. \ref{e}).
\end{description}

We also investigated the charge-orbital model obtained as a result of full spin-polarization. The model can be mapped to a Hubbard model with effective interaction $U_\text{eff} =  U-3J$. For repulsive side, we find the FM+ODW phase with massless charge and orbital modes is stable at most of the fillings. When two bands are equivalent, we have a charge gap at $n=1$, while we expect an orbital selective Mott phase around $n=1$ for inequivalent bands. The attractive side is dominated by orbital singlet superconductivity with an orbital gap when two bands are equal. On the other hand, for inequivalent bands, orbital singlet and triplet superconductivity have the longest correlation and the orbital gap is absent. The charge gap does not develop in the attractive side.

Finally we refer to possible experimental consequences. From the band calculation, it is plausible that the real system exhibits a ferromagnetic state, meaning that the system has relatively strong correlations.\cite{Zaki} In particular, since $U$ is much larger than $J$ in real materials, we guess the best possible ground state is a ferromagnetic orbital density wave from the discussion in Section \ref{strong}. Of course, we have to note that the result is based on strong assumptions such that the system is fully spin polarized, and that the number of the bands is just two.

\begin{acknowledgements}
We thank R. Osgood, and N. Zaki for fruitful discussion and the data they shared with us. This work was supported by the Department of Energy Contract Nos. DE-FG02-04ER46157 (JO), and DE-FG02-04ER46169 (AJM).
\end{acknowledgements}

\appendix
\section{Abelian Bosonization}
\label{bos}
In this appendix, we briefly recall formulas from Abelian-bosonization which are needed in the text.\cite{Voit1995a,giamarchi2004quantum, gogolin2004bosonization} The following formula gives transformation from a fermionic Hamiltonian to a bosonic Hamiltonian
\begin{equation}
\psi_{m \sigma r } = \frac{\eta_{m \sigma}}{\sqrt{2\pi \alpha}} e^{\mp i  \Phi_{m \sigma r} }
\end{equation}
where $m = A, B$ is orbital, and $r = R, L$ is chirality. The bosonic fields satisfy the commutation relations,
\begin{equation}
\begin{split}
[\Phi _{m \sigma R(L) }(x) , \Phi_{m' \sigma' R(L)} (x')]& = \pm  i \pi \delta_{m m'}\delta_{\sigma \sigma'} \text{sgn} (x-x')\\
[\Phi _{m \sigma R }(x) , \Phi_{m' \sigma' L} (x')] &= i \pi \delta_{m m'}\delta_{\sigma \sigma'}.
\end{split}
\end{equation}
The Majorana fermions take care of the fermionic properties and obey anti-commutation relations
\begin{equation}
\{ \eta_{m \sigma}, \eta_{m' \sigma'} \} = 2 \delta_{m m'}\delta_{\sigma \sigma'}.
\end{equation}
A more convenient representation is given by the non-chiral fields,
\begin{equation}
\phi_{m \sigma}, \theta_{m \sigma}  = \frac{1}{2}( \Phi_{m \sigma L} \pm \Phi_{m \sigma R}).
\end{equation}
They are connected to density and current as $\nabla \phi \propto n$ and $\nabla \theta \propto j$, and satisfy commutation relations,
\begin{equation}
\begin{split}
[\phi_{m \sigma} (x), \phi_{m' \sigma'}(x') ] &=[\theta_{m \sigma} (x), \theta_{m' \sigma'}(x') ] = 0\\
[\phi_{m \sigma} (x), \theta_{m' \sigma'}(x') ] &= -i\pi  \delta_{m m'}\delta_{\sigma \sigma'} \Theta (x'-x).
\end{split}
\end{equation}
Finally we move to different combinations of these fields,
\begin{equation}
\begin{bmatrix}
 \phi_{c0}\\ 
 \phi_{c\pi} \\ 
 \phi_{s0} \\ 
\phi_{s\pi}
\end{bmatrix} 
= \frac{1}{2}
\begin{bmatrix}
1 & 1 &1  &1  \\ 
 1& 1 &-1  &-1  \\ 
 1& -1 &1  &-1  \\ 
 1& -1 &-1  &1 
\end{bmatrix} 
\begin{bmatrix}
 \phi_{A\uparrow}\\ 
 \phi_{A\downarrow} \\ 
 \phi_{B\uparrow} \\ 
\phi_{B\downarrow}
\end{bmatrix} ,
\end{equation}
where $\mu = (c, s)$ represents charge or spin modes, and $\nu = (0, \pi)$ give bonding/antibonding basis. $\theta$'s are transformed in the same manner.

The sign of each coupling constant is determined by Klein factors, and by a commutator between different chirality, $[ \Phi_{R,n}(x), \Phi_{L,n'}(x') ] = i\pi \delta_{n,n'}$. The eigenvalues of Klein factors composed of two Majorana fermions (different from the ones introduced for refermionization) are taken to be 
\begin{multline}
i=\eta_{As}\eta_{Bs} = \eta_{A\uparrow}\eta_{A\downarrow} \\
=\eta_{A\uparrow}\eta_{B\downarrow} = \eta_{B\uparrow}\eta_{A\downarrow} = -\eta_{B\uparrow}\eta_{B\downarrow}.
\end{multline}

\section{RG equations for unequal velocities}
\label{bf}
When two Fermi velocities are different, it is more convenient to use current operators than using refermionization. We follow the notation of Ref. \onlinecite{Balents1996} with slight modification:
\begin{align}
J_{m r} = \sum _{s s'} \psi^{\dagger}_{m s r} \psi_{m s' r},  & \ \  \boldsymbol{J}_{m r} = \sum _{s s'} \psi^{\dagger}_{m s  r} \boldsymbol{\sigma}_{s s'} \psi_{m s' r}\\
L_{r} = \sum _{s s'} \psi^{\dagger}_{A s r} \psi_{B s' r}, & \ \  \boldsymbol{L}_{r} = \sum _{s s'} \psi^{\dagger}_{A s  r} \boldsymbol{\sigma}_{s s'} \psi_{B s' r}\\
M_{mr} = -i \psi_{m \uparrow r}\psi_{m \downarrow r}, & \ \  N_{r s s'} = \psi_{r A s}\psi_{r B s'}.
\end{align}
When $k_A \neq k_B$, the interactions terms are given by,
\begin{equation}
\begin{split}
-\mathcal{H}_\text{int} &=  \widetilde{g}_{1\rho}J_{AR}J_{AL} + \widetilde{g}_{1\sigma}\boldsymbol{J}_{AR}\cdot \boldsymbol{J}_{AL}\\
&+ \widetilde{g}_{2\rho}J_{BR}J_{BL} + \widetilde{g}_{2\sigma}\boldsymbol{J}_{BR}\cdot\boldsymbol{J}_{BL}\\
&+\widetilde{g}_{x\rho}\left( J_{AR}J_{BL}+ J_{BR}J_{AL}\right)\\
&+ \widetilde{g}_{x\sigma}\left( \boldsymbol{J}_{AR}\cdot\boldsymbol{J}_{BL}+\boldsymbol{J}_{BR}\cdot\boldsymbol{J}_{AL}\right)\\
&+ \widetilde{g}_{t\rho} ( L_{R} L_{L}+ \text{h.c.} ) + \widetilde{g}_{t\sigma} ( \boldsymbol{L}_{R} \cdot \boldsymbol{L}_{L} + \text{h.c.}).
\end{split}
\end{equation}
This expression is formally the same as the one given in Ref. \onlinecite{Balents1996}. When $k_{A} =k_{B}$, we have additional processes,
\begin{equation}
-\mathcal{H}'_\text{int} = \widetilde{g}_{a\rho}\left( L_{R} L_{L}^{\dagger} + \text{h.c.}\right) + \widetilde{g}_{a\sigma}\left(\boldsymbol{L}_{R} \cdot \boldsymbol{L}_{L}^{\dagger} + \text{h.c.}\right).
\end{equation}
Umklapp processes are allowed when the filling is commensurate ($n=2$):
\begin{equation}
\begin{split}
-\mathcal{H}''_\text{int} &=  \widetilde{g}_{1u}\left( M_{AR} M_{AL}^{\dagger}+ \text{h.c.}\right) \\
&+ \widetilde{g}_{2u}\left( M_{BR} M_{BL}^{\dagger} + \text{h.c.}\right)\\
&+ \widetilde{g}_{xu}\left( M_{AR} M_{BL}^{\dagger} + M_{BR} M_{AL}^{\dagger} +\text{h.c.}\right)\\
&+\widetilde{g}_{tu\rho}\left( N_{R \alpha \beta}^{\dagger} N_{L \alpha \beta} - N_{R \alpha \beta}^{\dagger} N_{L \beta \alpha} +   \text{h.c.}\right)\\
&+ \widetilde{g}_{tu\sigma}\left( N_{R \alpha \beta}^{\dagger} N_{L \alpha \beta} + N_{R \alpha \beta}^{\dagger} N_{L \beta \alpha}+   \text{h.c.}\right).
\end{split}
\end{equation}
The $\widetilde{g}_{1u}$ and $\widetilde{g}_{2u}$ processes are allowed only when each band has commensurate filling, i.e., $k_{m} = \pi/2$. We ignore all the chiral scattering processes, since they only renormalize the velocities. 

In the following, we use the renormalized coupling constants, $y_i = \widetilde{g_{i}} \pi^{-1} (v_{A} + v_{B})^{-1} $. The RG equations for the $k_{A} = k_{B} = \pi/2$ case are
\begin{widetext}
\begin{equation}
\begin{split}
\dot{y}_{1\rho} &= -\beta  \left(y_{a\rho}^2+3 y_{a\sigma}^2+3 y_{tu\sigma}^2+y_{tu\rho}^2-y_{t\rho}^2-3 y_{t\sigma}^2\right)-\alpha  y_{1u}^2\\
\dot{y}_{2\rho} &=-\alpha  \left(y_{a\rho}^2+3 y_{a\sigma}^2+3 y_{tu\sigma}^2+y_{tu\rho}^2-y_{t\rho}^2-3 y_{t\sigma}^2\right)-\beta  y_{2 u}^2\\
\dot{y}_{x\rho} &=y_{a\rho}^2+3 y_{a\sigma}^2-3 y_{tu\sigma}^2-y_{tu\rho}^2-y_{t\rho}^2-3 y_{t\sigma}^2-y_{xu}^2
\end{split}
\end{equation}
\begin{equation}
\begin{split}
\dot{y}_{1\sigma} &=-2 \beta  \left(y_{a\sigma} \left(y_{a\sigma} +y_{a\rho}\right)+y_{tu\sigma} \left(y_{tu\sigma}+y_{tu\rho}\right)+y_{t\sigma} \left(y_{t\sigma}-y_{t\rho}\right) \right)-4 \alpha  y_{1\sigma}^2\\
\dot{y}_{2\sigma} &=-2 \alpha  \left(y_{a\sigma} \left(y_{a\sigma} + y_{a\rho}\right)+y_{tu\sigma} \left(y_{tu\sigma}+y_{tu\rho}\right)+y_{t\sigma} \left(y_{t\sigma}-y_{t\rho}\right)\right)-4 \beta  y_{2 \sigma }^2\\
\dot{y}_{x\sigma} &=-2 \left(y_{a\sigma} \left(y_{a\sigma}-y_{a\rho}\right)+y_{tu\sigma} \left(y_{tu\sigma}-y_{tu\rho}\right)+y_{t\sigma} \left(y_{t\sigma}+y_{t\rho}\right)\right)-4 y_{x\sigma}^2
\end{split}
\end{equation}
\begin{equation}
\begin{split}
\dot{y}_{t\rho} &=-2 y_{tu\rho} y_{xu}+y_{t\rho} y_{c-}+3 y_{t\sigma}y_{s-}\\
\dot{y}_{t\sigma} &= 2 y_{tu\sigma} y_{xu}+y_{t\rho} y_{s-} +y_{t\sigma} \left(y_{c-} -2y_{s+}\right)
\end{split}
\end{equation}
\begin{equation}
\begin{split}
\dot{y}_{a\rho} &= -y_{tu\rho} \left(\alpha  y_{1u}+\beta  y_{2 u}\right)- y_{a\rho}y_{c-} - 3 y_{a\sigma}y_{s-} \\
\dot{y}_{a \sigma} &= - y_{tu\sigma} \left(\alpha  y_{1u}+\beta  y_{2 u}\right)- y_{a\rho} y_{s-}-  y_{a\sigma} \left(y_{c-}+2y_{s+} \right)
\end{split}
\end{equation}
\begin{equation}
\begin{split}
\dot{y}_{1u} &=-4 \left(3 \beta  y_{a\sigma} y_{tu\sigma}+\beta  y_{a\rho} y_{tu\rho}+\alpha  y_{1u} y_{1\rho}\right)\\
\dot{y}_{2u} &=-4 \left(3 \alpha  y_{a\sigma} y_{tu\sigma}+\alpha  y_{a\rho} y_{tu\rho}+\beta  y_{2 u} y_{2 \rho }\right)\\
\dot{y}_{xu} &=4 \left(3 y_{t\sigma} y_{tu\sigma} - y_{t\rho}y_{tu\rho} -y_{xu} y_{x\rho}\right)
\end{split}
\end{equation}
\begin{equation}
\begin{split}
\dot{y}_{tu\rho} &=-y_{a\rho} \left(\alpha  y_{1u}+\beta  y_{2 u}\right)-2 y_{t\rho} y_{xu}-y_{tu\rho} y_{c+}-3 y_{tu\sigma} y_{s-}\\
\dot{y}_{tu\sigma} &=-y_{a\sigma} \left(\alpha  y_{1u}+\beta  y_{2 u}\right)+2 y_{t\sigma} y_{xu}-y_{tu\rho} y_{s-}-y_{tu\sigma} \left(y_{c+} + 2y_{s+}\right),
\end{split}
\end{equation}
\end{widetext}
where we defined $y_{c(s)\pm} = \alpha y_{1\rho(\sigma)} + \beta y_{2\rho(\sigma)} \pm2y_{x\rho(\sigma)} $ with $\alpha = (v_{A} +v_{B})/(2v_{A})$, and $\beta = (v_{A} +v_{B})/(2v_{B})$. For doped cases, and $k_{A} \neq k_{B}$ cases, the coupling constants which are not allowed by momentum conservation should be removed.

As we mentioned, the asymptotic behavior of a RG flow is captured by the ansatz \eqref{ray}, and now the ratios of coupling constants at fixed points depend on velocity differences. However, we can easily distinguish phases with different fixed point structure by looking at the signs of relevant couplings, and irrelevant couplings. In that sense, we identify phases as the same ones when the relevant couplings and the signs are the same. When the relevant couplings are different, or the signs of renormalized couplings are different, we regard them as different phases.


\newpage
\begin{thebibliography}{100}
\expandafter\ifx\csname natexlab\endcsname\relax\def\natexlab#1{#1}\fi
\expandafter\ifx\csname bibnamefont\endcsname\relax
  \def\bibnamefont#1{#1}\fi
\expandafter\ifx\csname bibfnamefont\endcsname\relax
  \def\bibfnamefont#1{#1}\fi
\expandafter\ifx\csname citenamefont\endcsname\relax
  \def\citenamefont#1{#1}\fi
\expandafter\ifx\csname url\endcsname\relax
  \def\url#1{\texttt{#1}}\fi
\expandafter\ifx\csname urlprefix\endcsname\relax\def\urlprefix{URL }\fi
\providecommand{\bibinfo}[2]{#2}
\providecommand{\eprint}[2][]{\url{#2}}


\bibitem[{\citenamefont{Varma and Zawadowski}(1985)}]{Varma1985a}
\bibinfo{author}{\bibfnamefont{C.~M.}~\bibnamefont{Varma}} \bibnamefont{and}
  \bibinfo{author}{\bibfnamefont{A.}~\bibnamefont{Zawadowski}},
  \bibinfo{journal}{Phys. Rev. B} \textbf{\bibinfo{volume}{32}},
  \bibinfo{pages}{7399} (\bibinfo{year}{1985}).
  
\bibitem[{\citenamefont{Strong and Millis}(1994)}]{Strong1994a}
\bibinfo{author}{\bibfnamefont{S.~P.}~\bibnamefont{Strong}} \bibnamefont{and}
  \bibinfo{author}{\bibfnamefont{A.~J.}~\bibnamefont{Millis}},
  \bibinfo{journal}{Phys. Rev. B} \textbf{\bibinfo{volume}{50}},
  \bibinfo{pages}{9911} (\bibinfo{year}{1994}).

\bibitem[{\citenamefont{Fujimoto and Kawakami}(1994)}]{Fujimoto1994}
\bibinfo{author}{\bibfnamefont{S.}~\bibnamefont{Fujimoto}} \bibnamefont{and}
  \bibinfo{author}{\bibfnamefont{N.}~\bibnamefont{Kawakami}},
  \bibinfo{journal}{J. Phys. Soc. Jpn.}
  \textbf{\bibinfo{volume}{63}}, \bibinfo{pages}{4322} (\bibinfo{year}{1994}).

\bibitem[{\citenamefont{Fabrizio et~al.}(1992)\citenamefont{Fabrizio, Parola,
  and Tosatti}}]{Fabrizio1992}
\bibinfo{author}{\bibfnamefont{M.}~\bibnamefont{Fabrizio}},
  \bibinfo{author}{\bibfnamefont{A.}~\bibnamefont{Parola}}, \bibnamefont{and}
  \bibinfo{author}{\bibfnamefont{E.}~\bibnamefont{Tosatti}},
  \bibinfo{journal}{Phys. Rev. B} \textbf{\bibinfo{volume}{46}},
  \bibinfo{pages}{3159} (\bibinfo{year}{1992}).

\bibitem[{\citenamefont{Finkel’stein and Larkin}(1993)}]{Finkelstein1993}
\bibinfo{author}{\bibfnamefont{A.}~\bibnamefont{Finkel’stein}}
  \bibnamefont{and} \bibinfo{author}{\bibfnamefont{A.}~\bibnamefont{Larkin}},
  \bibinfo{journal}{Phys. Rev. B} \textbf{\bibinfo{volume}{47}},
  \bibinfo{pages}{10461} (\bibinfo{year}{1993}).

\bibitem[{\citenamefont{Khveshchenko and Rice}(1994)}]{Khveshchenko1994d}
\bibinfo{author}{\bibfnamefont{D.}~\bibnamefont{Khveshchenko}}
  \bibnamefont{and} \bibinfo{author}{\bibfnamefont{T.~M.}~\bibnamefont{Rice}},
  \bibinfo{journal}{Phys. Rev. B} \textbf{\bibinfo{volume}{50}},
  \bibinfo{pages}{252} (\bibinfo{year}{1994}).

\bibitem[{\citenamefont{Balents and Fisher}(1996)}]{Balents1996}
\bibinfo{author}{\bibfnamefont{L.}~\bibnamefont{Balents}} \bibnamefont{and}
  \bibinfo{author}{\bibfnamefont{M.~P.~A.}~\bibnamefont{Fisher}},
  \bibinfo{journal}{Phys. Rev. B}
  \textbf{\bibinfo{volume}{53}}, \bibinfo{pages}{12133} (\bibinfo{year}{1996}).

\bibitem[{\citenamefont{Schulz}(1996)}]{Schulz1996}
\bibinfo{author}{\bibfnamefont{H.~J.}~\bibnamefont{Schulz}},
  \bibinfo{journal}{Phys. Rev. B} \textbf{\bibinfo{volume}{53}},
  \bibinfo{pages}{2959} (\bibinfo{year}{1996}).

\bibitem[{\citenamefont{Shelton and Tsvelik}(1996)}]{Shelton1996a}
\bibinfo{author}{\bibfnamefont{D.}~\bibnamefont{Shelton}} \bibnamefont{and}
  \bibinfo{author}{\bibfnamefont{A.~M.}~\bibnamefont{Tsvelik}},
  \bibinfo{journal}{Phys. Rev. B} \textbf{\bibinfo{volume}{53}},
  \bibinfo{pages}{14036} (\bibinfo{year}{1996}).

\bibitem[{\citenamefont{Lin et~al.}(1998)\citenamefont{Lin, Balents, and
  Fisher}}]{Lin1998}
\bibinfo{author}{\bibfnamefont{H.~H.} \bibnamefont{Lin}},
  \bibinfo{author}{\bibfnamefont{L.}~\bibnamefont{Balents}}, \bibnamefont{and}
  \bibinfo{author}{\bibfnamefont{M.~P.~A.}~\bibnamefont{Fisher}},
  \bibinfo{journal}{Phys. Rev. B} \textbf{\bibinfo{volume}{58}},
  \bibinfo{pages}{1794} (\bibinfo{year}{1998}).
  
\bibitem[{\citenamefont{Lee et~al.}(2005)\citenamefont{Lee, Marston, and
  Fjaerestad}}]{Lee2005}
\bibinfo{author}{\bibfnamefont{S.}~\bibnamefont{Lee}},
  \bibinfo{author}{\bibfnamefont{J.~B.}~\bibnamefont{Marston}}, \bibnamefont{and}
  \bibinfo{author}{\bibfnamefont{J.~O.}~\bibnamefont{Fjaerestad}},
  \bibinfo{journal}{Phys. Rev. B} \textbf{\bibinfo{volume}{72}},
  \bibinfo{pages}{075126} (\bibinfo{year}{2005}).


\bibitem[{\citenamefont{Chudzinski et~al.}(2008)\citenamefont{Chudzinski,
  Gabay, and Giamarchi}}]{Chudzinski2008}
\bibinfo{author}{\bibfnamefont{P.}~\bibnamefont{Chudzinski}},
  \bibinfo{author}{\bibfnamefont{M.}~\bibnamefont{Gabay}}, \bibnamefont{and}
  \bibinfo{author}{\bibfnamefont{T.}~\bibnamefont{Giamarchi}},
  \bibinfo{journal}{Phys. Rev. B} \textbf{\bibinfo{volume}{78}},
  , \bibinfo{pages}{075124} (\bibinfo{year}{2008}).

\bibitem[{\citenamefont{Shelton et~al.}(1996)\citenamefont{Shelton, Nersesyan,
  and Tsvelik}}]{Shelton1996b}
\bibinfo{author}{\bibfnamefont{D.~G.}~\bibnamefont{Shelton}},
  \bibinfo{author}{\bibfnamefont{A.~A.}~\bibnamefont{Nersesyan}},
  \bibnamefont{and} \bibinfo{author}{\bibfnamefont{A.}~\bibnamefont{Tsvelik}},
  \bibinfo{journal}{Phys. Rev. B}
  \textbf{\bibinfo{volume}{53}}, \bibinfo{pages}{8521} (\bibinfo{year}{1996}).

\bibitem[{\citenamefont{Kim et~al.}(2000)\citenamefont{Kim, F\'{a}th, and
  S\'{o}lyom}}]{Kim2000}
\bibinfo{author}{\bibfnamefont{E.}~\bibnamefont{Kim}},
  \bibinfo{author}{\bibfnamefont{G.}~\bibnamefont{F\'{a}th}}, \bibnamefont{and}
  \bibinfo{author}{\bibfnamefont{J.}~\bibnamefont{S\'{o}lyom}},
  \bibinfo{journal}{Phys. Rev. B} \textbf{\bibinfo{volume}{62}},
  \bibinfo{pages}{965} (\bibinfo{year}{2000}).

\bibitem[{\citenamefont{Azaria et~al.}(1999)\citenamefont{Azaria, Gogolin,
  Lecheminant, and Nersesyan}}]{Azaria1999}
\bibinfo{author}{\bibfnamefont{P.}~\bibnamefont{Azaria}},
  \bibinfo{author}{\bibfnamefont{A.}~\bibnamefont{Gogolin}},
  \bibinfo{author}{\bibfnamefont{P.}~\bibnamefont{Lecheminant}},
  \bibnamefont{and}
  \bibinfo{author}{\bibfnamefont{A.~A.}~\bibnamefont{Nersesyan}},
  \bibinfo{journal}{Phys. Rev. Lett.} \textbf{\bibinfo{volume}{83}},
  \bibinfo{pages}{624} (\bibinfo{year}{1999}).

\bibitem[{\citenamefont{Tsuchiizu and Furusaki}(2002)}]{Tsuchiizu2002}
\bibinfo{author}{\bibfnamefont{M.}~\bibnamefont{Tsuchiizu}} \bibnamefont{and}
  \bibinfo{author}{\bibfnamefont{A.}~\bibnamefont{Furusaki}},
  \bibinfo{journal}{Phys. Rev. B} \textbf{\bibinfo{volume}{66}},
   \bibinfo{pages}{245106} (\bibinfo{year}{2002}).

\bibitem[{\citenamefont{Wu et~al.}(2003)\citenamefont{Wu, Liu, and
  Fradkin}}]{Wu2003}
\bibinfo{author}{\bibfnamefont{C.}~\bibnamefont{Wu}},
  \bibinfo{author}{\bibfnamefont{W.~V.}~\bibnamefont{Liu}},
  \bibnamefont{and} \bibinfo{author}{\bibfnamefont{E.}~\bibnamefont{Fradkin}},
  \bibinfo{journal}{Phys. Rev. B} \textbf{\bibinfo{volume}{68}},
  \bibinfo{pages}{115104} (\bibinfo{year}{2003}).

\bibitem[{\citenamefont{Wang et~al.}(2008)\citenamefont{Wang, Yilmaz, Knox,
  Zaki, Dadap, Valla, Johnson, and Osgood}}]{Wang2008}
\bibinfo{author}{\bibfnamefont{S.} \bibnamefont{Wang}},
  \bibinfo{author}{\bibfnamefont{M.~B.}~\bibnamefont{Yilmaz}},
  \bibinfo{author}{\bibfnamefont{K.~R.}~\bibnamefont{Knox}},
  \bibinfo{author}{\bibfnamefont{N.}~\bibnamefont{Zaki}},
  \bibinfo{author}{\bibfnamefont{J.~I.}~\bibnamefont{Dadap}},
  \bibinfo{author}{\bibfnamefont{R.~M.}~\bibnamefont{Osgood}},
  \bibinfo{author}{\bibfnamefont{T.}~\bibnamefont{Valla}}, \bibnamefont{and}
  \bibinfo{author}{\bibfnamefont{P.~D.}~\bibnamefont{Johnson}}, 
  \bibinfo{journal}{Phys. Rev. B} \textbf{\bibinfo{volume}{77}},
  \bibinfo{pages}{115448} (\bibinfo{year}{2008}).

\bibitem[{\citenamefont{Zaki et~al.}(2009)\citenamefont{Zaki, Potapenko,
  Johnson, and Osgood}}]{Zaki2009}
\bibinfo{author}{\bibfnamefont{N.}~\bibnamefont{Zaki}},
  \bibinfo{author}{\bibfnamefont{D.}~\bibnamefont{Potapenko}},
  \bibinfo{author}{\bibfnamefont{P.~D.}~\bibnamefont{Johnson}}, \bibnamefont{and}
  \bibinfo{author}{\bibfnamefont{R.~M.}~\bibnamefont{Osgood}},
  \bibinfo{journal}{Phys. Rev. B} \textbf{\bibinfo{volume}{80}},
  \bibinfo{pages}{155419} (\bibinfo{year}{2009}).

\bibitem[{\citenamefont{Konik et~al.}(2002)\citenamefont{Konik, Saleur, and
  Ludwig}}]{Konik2002}
\bibinfo{author}{\bibfnamefont{R.}~\bibnamefont{Konik}},
  \bibinfo{author}{\bibfnamefont{H.}~\bibnamefont{Saleur}}, \bibnamefont{and}
  \bibinfo{author}{\bibfnamefont{A.}~\bibnamefont{Ludwig}},
  \bibinfo{journal}{Phys. Rev. B} \textbf{\bibinfo{volume}{66}},
  \bibinfo{pages}{075105} (\bibinfo{year}{2002}).

\bibitem[{\citenamefont{Controzzi and Tsvelik}(2005)}]{Controzzi2005}
\bibinfo{author}{\bibfnamefont{D.}~\bibnamefont{Controzzi}} \bibnamefont{and}
  \bibinfo{author}{\bibfnamefont{A.~M.}~\bibnamefont{Tsvelik}},
  \bibinfo{journal}{Phys. Rev. B} \textbf{\bibinfo{volume}{72}},
  \bibinfo{pages}{035110} (\bibinfo{year}{2005}).

\bibitem[{\citenamefont{Boulat et~al.}(2009)\citenamefont{Boulat, Azaria, and
  Lecheminant}}]{Boulat2009}
\bibinfo{author}{\bibfnamefont{E.}~\bibnamefont{Boulat}},
  \bibinfo{author}{\bibfnamefont{P.}~\bibnamefont{Azaria}}, \bibnamefont{and}
  \bibinfo{author}{\bibfnamefont{P.}~\bibnamefont{Lecheminant}},
  \bibinfo{journal}{Nuclear Physics B} \textbf{\bibinfo{volume}{822}},
  \bibinfo{pages}{367} (\bibinfo{year}{2009}).

\bibitem[{\citenamefont{Nonne et~al.}(2010)\citenamefont{Nonne, Boulat,
  Capponi, and Lecheminant}}]{Nonne2010}
\bibinfo{author}{\bibfnamefont{H.}~\bibnamefont{Nonne}},
  \bibinfo{author}{\bibfnamefont{E.}~\bibnamefont{Boulat}},
  \bibinfo{author}{\bibfnamefont{S.}~\bibnamefont{Capponi}}, \bibnamefont{and}
  \bibinfo{author}{\bibfnamefont{P.}~\bibnamefont{Lecheminant}},
  \bibinfo{journal}{Phys. Rev. B} \textbf{\bibinfo{volume}{82}},
  \bibinfo{pages}{155134} (\bibinfo{year}{2010}).

\bibitem[{\citenamefont{Kramers and Wannier}(1941)}]{Kramers1941}
\bibinfo{author}{\bibfnamefont{H.}~\bibnamefont{Kramers}} \bibnamefont{and}
  \bibinfo{author}{\bibfnamefont{G.}~\bibnamefont{Wannier}},
  \bibinfo{journal}{Phys. Rev.} \textbf{\bibinfo{volume}{60}},
  \bibinfo{pages}{252} (\bibinfo{year}{1941}).

\bibitem[{\citenamefont{Tsvelik}(2011)}]{Tsvelik2011}
\bibinfo{author}{\bibfnamefont{A.~M.}~\bibnamefont{Tsvelik}},
  \bibinfo{journal}{Phys. Rev. B} \textbf{\bibinfo{volume}{83}},
  \bibinfo{pages}{104405} (\bibinfo{year}{2011}).

\bibitem[{\citenamefont{Gambardella et~al.}(2002)\citenamefont{Gambardella,
  Dallmeyer, Maiti, Malagoli, Eberhardt, Kern, and Carbone}}]{Gambardella2002a}
\bibinfo{author}{\bibfnamefont{P.}~\bibnamefont{Gambardella}},
  \bibinfo{author}{\bibfnamefont{A.}~\bibnamefont{Dallmeyer}},
  \bibinfo{author}{\bibfnamefont{K.}~\bibnamefont{Maiti}},
  \bibinfo{author}{\bibfnamefont{M.~C.} \bibnamefont{Malagoli}},
  \bibinfo{author}{\bibfnamefont{W.}~\bibnamefont{Eberhardt}},
  \bibinfo{author}{\bibfnamefont{K.}~\bibnamefont{Kern}}, \bibnamefont{and}
  \bibinfo{author}{\bibfnamefont{C.}~\bibnamefont{Carbone}},
  \bibinfo{journal}{Nature} \textbf{\bibinfo{volume}{416}},
  \bibinfo{pages}{301} (\bibinfo{year}{2002}).

\bibitem[{\citenamefont{Koga et~al.}(2004)\citenamefont{Koga, Kawakami, Rice,
  and Sigrist}}]{Koga2004}
\bibinfo{author}{\bibfnamefont{A.}~\bibnamefont{Koga}},
  \bibinfo{author}{\bibfnamefont{N.}~\bibnamefont{Kawakami}},
  \bibinfo{author}{\bibfnamefont{T.~M.}~\bibnamefont{Rice}}, \bibnamefont{and}
  \bibinfo{author}{\bibfnamefont{M.}~\bibnamefont{Sigrist}},
  \bibinfo{journal}{Phys. Rev. Lett.} \textbf{\bibinfo{volume}{92}},
  \bibinfo{pages}{216402} (\bibinfo{year}{2004}).

\bibitem[{\citenamefont{Okamoto and Millis}(2011)}]{Okamoto2011}
\bibinfo{author}{\bibfnamefont{J.} \bibnamefont{Okamoto}} \bibnamefont{and}
  \bibinfo{author}{\bibfnamefont{A.~J.} \bibnamefont{Millis}}, \bibinfo{journal}{Arxiv
  preprint arXiv:1108.1389} (\bibinfo{year}{2011}).

\bibitem[{\citenamefont{Lee et~al.}(2004)\citenamefont{Lee, Azaria, and
  Boulat}}]{Lee2004}
\bibinfo{author}{\bibfnamefont{H.}~\bibnamefont{Lee}},
  \bibinfo{author}{\bibfnamefont{P.}~\bibnamefont{Azaria}}, \bibnamefont{and}
  \bibinfo{author}{\bibfnamefont{E.}~\bibnamefont{Boulat}},
  \bibinfo{journal}{Phys. Rev. B} \textbf{\bibinfo{volume}{69}},
  \bibinfo{pages}{155109} (\bibinfo{year}{2004}).

\bibitem[{\citenamefont{Voit}(1995)}]{Voit1995a}
\bibinfo{author}{\bibfnamefont{J.}~\bibnamefont{Voit}},
  \bibinfo{journal}{Rep. Prog. Phys.} \textbf{\bibinfo{volume}{58}},
  \bibinfo{pages}{977} (\bibinfo{year}{1995}).
  
\bibitem[{\citenamefont{Giamarchi}(2004)}]{giamarchi2004quantum}
\bibinfo{author}{\bibfnamefont{T.}~\bibnamefont{Giamarchi}},
  \emph{\bibinfo{title}{{Quantum physics in one dimension}}} (\bibinfo{publisher}{Clarendon},
  \bibinfo{year}{2004}).

\bibitem[{\citenamefont{Gogolin et~al.}(2004)\citenamefont{Gogolin, Nersesyan,
  and Tsvelik}}]{gogolin2004bosonization}
\bibinfo{author}{\bibfnamefont{A.~O.} \bibnamefont{Gogolin}},
  \bibinfo{author}{\bibfnamefont{A.~A.} \bibnamefont{Nersesyan}},
  \bibnamefont{and} \bibinfo{author}{\bibfnamefont{A.~M.}
  \bibnamefont{Tsvelik}}, \emph{\bibinfo{title}{{Bosonization and strongly
  correlated systems}}} (\bibinfo{publisher}{Cambridge University Press},
  \bibinfo{year}{2004}).

\bibitem[{\citenamefont{Nishiyama et~al.}(1995)\citenamefont{Nishiyama, Hatano,
  and Suzuki}}]{Nishiyama1995}
\bibinfo{author}{\bibfnamefont{Y.}~\bibnamefont{Nishiyama}},
  \bibinfo{author}{\bibfnamefont{N.}~\bibnamefont{Hatano}}, \bibnamefont{and}
  \bibinfo{author}{\bibfnamefont{M.}~\bibnamefont{Suzuki}},
  \bibinfo{journal}{J. Phys. Soc. Jpn.}
  \textbf{\bibinfo{volume}{64}}, \bibinfo{pages}{1967} (\bibinfo{year}{1995}).

\bibitem[{\citenamefont{Affleck et~al.}(1987)\citenamefont{Affleck, Kennedy,
  Lieb, and Tasaki}}]{Affleck1987}
\bibinfo{author}{\bibfnamefont{I.}~\bibnamefont{Affleck}},
  \bibinfo{author}{\bibfnamefont{T.}~\bibnamefont{Kennedy}},
  \bibinfo{author}{\bibfnamefont{E.}~\bibnamefont{Lieb}}, \bibnamefont{and}
  \bibinfo{author}{\bibfnamefont{H.}~\bibnamefont{Tasaki}},
  \bibinfo{journal}{Phys. Rev. Lett.} \textbf{\bibinfo{volume}{59}},
  \bibinfo{pages}{799} (\bibinfo{year}{1987}).

\bibitem[{\citenamefont{Berg et~al.}(2008)\citenamefont{Berg, {Dalla Torre},
  Giamarchi, and Altman}}]{Berg2008}
\bibinfo{author}{\bibfnamefont{E.}~\bibnamefont{Berg}},
  \bibinfo{author}{\bibfnamefont{E.}~\bibnamefont{{Dalla Torre}}},
  \bibinfo{author}{\bibfnamefont{T.}~\bibnamefont{Giamarchi}},
  \bibnamefont{and} \bibinfo{author}{\bibfnamefont{E.}~\bibnamefont{Altman}},
  \bibinfo{journal}{Phys. Rev. B} \textbf{\bibinfo{volume}{77}},
  \bibinfo{pages}{245119} (\bibinfo{year}{2008}).

\bibitem[{\citenamefont{Assaraf et~al.}(2004)\citenamefont{Assaraf, Azaria,
  Boulat, Caffarel, and Lecheminant}}]{Assaraf2004}
\bibinfo{author}{\bibfnamefont{R.}~\bibnamefont{Assaraf}},
  \bibinfo{author}{\bibfnamefont{P.}~\bibnamefont{Azaria}},
  \bibinfo{author}{\bibfnamefont{E.}~\bibnamefont{Boulat}},
  \bibinfo{author}{\bibfnamefont{M.}~\bibnamefont{Caffarel}}, \bibnamefont{and}
  \bibinfo{author}{\bibfnamefont{P.}~\bibnamefont{Lecheminant}},
  \bibinfo{journal}{Phys. Rev. Lett.} \textbf{\bibinfo{volume}{93}},
  \bibinfo{pages}{16407} (\bibinfo{year}{2004}).

\bibitem[{\citenamefont{Bunder and Lin}(2007)}]{Bunder2007}
\bibinfo{author}{\bibfnamefont{J.}~\bibnamefont{Bunder}} \bibnamefont{and}
  \bibinfo{author}{\bibfnamefont{H.~H.} \bibnamefont{Lin}},
  \bibinfo{journal}{Phys. Rev. B} \textbf{\bibinfo{volume}{75}},
  \bibinfo{pages}{075418} (\bibinfo{year}{2007}).

\bibitem[{Note1()}]{Note1}
\bibinfo{note}{To get the same form, we have to redefine $\psi ^{c\pi
  }_{L} =   \left ( \xi ^{5}_{L} - i \xi
  ^{6}_{L}\right )/\sqrt{2}$. Here the modes correspond to the new orbital part are $\xi
  _{5}$ and $\xi _{6}$, and the Ising mode is carried by $\xi _{4}$.}

\bibitem[{\citenamefont{Gross and Neveu}(1974)}]{Gross1974}
\bibinfo{author}{\bibfnamefont{D.}~\bibnamefont{Gross}} \bibnamefont{and}
  \bibinfo{author}{\bibfnamefont{A.}~\bibnamefont{Neveu}},
  \bibinfo{journal}{Phys. Rev. D} \textbf{\bibinfo{volume}{10}},
  \bibinfo{pages}{3235} (\bibinfo{year}{1974}).

\bibitem[{\citenamefont{Shankar}(1985)}]{Shankar1985}
\bibinfo{author}{\bibfnamefont{R.}~\bibnamefont{Shankar}},
  \bibinfo{journal}{Phys. Rev. Lett.} \textbf{\bibinfo{volume}{55}},
  \bibinfo{pages}{453} (\bibinfo{year}{1985}).

\bibitem[{\citenamefont{Cardy}(1996)}]{cardy1996scaling}
\bibinfo{author}{\bibfnamefont{J.~L.} \bibnamefont{Cardy}},
  \emph{\bibinfo{title}{{Scaling and renormalization in statistical physics}}} (\bibinfo{publisher}{Cambridge University Press}, \bibinfo{year}{1996}).

\bibitem[{\citenamefont{von Delft and Schoeller}(1998)}]{VonDelft1998}
\bibinfo{author}{\bibfnamefont{J.}~\bibnamefont{von Delft}} \bibnamefont{and}
  \bibinfo{author}{\bibfnamefont{H.}~\bibnamefont{Schoeller}},
  \bibinfo{journal}{Annalen der Physik} \textbf{\bibinfo{volume}{7}},
  \bibinfo{pages}{225} (\bibinfo{year}{1998}).

\bibitem[{\citenamefont{Chang et~al.}(2005)\citenamefont{Chang, Chen, and
  Lin}}]{Chang2005}
\bibinfo{author}{\bibfnamefont{M.}~\bibnamefont{Chang}},
  \bibinfo{author}{\bibfnamefont{W.}~\bibnamefont{Chen}}, \bibnamefont{and}
  \bibinfo{author}{\bibfnamefont{H.}~\bibnamefont{Lin}},
  \bibinfo{journal}{Prog. Theor. Phys. Suppl.}
  \textbf{\bibinfo{volume}{160}}, \bibinfo{pages}{79} (\bibinfo{year}{2005}).
  

\bibitem[{\citenamefont{Zamolodchikov}(1986)}]{Zamolodchikov1986a}
\bibinfo{author}{\bibfnamefont{A.}~\bibnamefont{Zamolodchikov}},
  \bibinfo{journal}{JETP Lett} \textbf{\bibinfo{volume}{43}},
  \bibinfo{pages}{9} (\bibinfo{year}{1986}).

\bibitem[{\citenamefont{Nakamura}(2000)}]{Nakamura2000}
\bibinfo{author}{\bibfnamefont{M.}~\bibnamefont{Nakamura}},
  \bibinfo{journal}{Phys. Rev. B}
  \textbf{\bibinfo{volume}{60}}, \bibinfo{pages}{16377} (\bibinfo{year}{2000}).

\bibitem[{\citenamefont{Tsuchiizu and Furusaki}(2002)}]{Tsuchiizu2002a}
\bibinfo{author}{\bibfnamefont{M.}~\bibnamefont{Tsuchiizu}} \bibnamefont{and}
  \bibinfo{author}{\bibfnamefont{A.}~\bibnamefont{Furusaki}},
  \bibinfo{journal}{Phys. Rev. Lett.} \textbf{\bibinfo{volume}{88}},
   \bibinfo{pages}{056402} (\bibinfo{year}{2002}).

\bibitem[{\citenamefont{Kugel and Khomskii}(1972)}]{Kugel1972}
\bibinfo{author}{\bibfnamefont{K.}~\bibnamefont{Kugel}} \bibnamefont{and}
  \bibinfo{author}{\bibfnamefont{D.}~\bibnamefont{Khomskii}},
  \bibinfo{journal}{JETP Lett.} \textbf{\bibinfo{volume}{15}}, \bibinfo{pages}{446} (\bibinfo{year}{1972}).

\bibitem[{\citenamefont{Cyrot and Lyon-Caen}(1975)}]{Cyrot1975}
\bibinfo{author}{\bibfnamefont{M.}~\bibnamefont{Cyrot}} \bibnamefont{and}
  \bibinfo{author}{\bibfnamefont{C.}~\bibnamefont{Lyon-Caen}},
  \bibinfo{journal}{J. Phys. (Paris)} \textbf{\bibinfo{volume}{36}},
  \bibinfo{pages}{253} (\bibinfo{year}{1975}).

\bibitem[{\citenamefont{Gill and Scalapino}(1987)}]{Gill1987}
\bibinfo{author}{\bibfnamefont{W.}~\bibnamefont{Gill}} \bibnamefont{and}
  \bibinfo{author}{\bibfnamefont{D.}~\bibnamefont{Scalapino}},
  \bibinfo{journal}{Phys. Rev. B} \textbf{\bibinfo{volume}{35}},
  \bibinfo{pages}{215} (\bibinfo{year}{1987}).

\bibitem[{\citenamefont{Sakamoto et~al.}(2002)\citenamefont{Sakamoto, Momoi,
  and Kubo}}]{Sakamoto2002}
\bibinfo{author}{\bibfnamefont{H.}~\bibnamefont{Sakamoto}},
  \bibinfo{author}{\bibfnamefont{T.}~\bibnamefont{Momoi}}, \bibnamefont{and}
  \bibinfo{author}{\bibfnamefont{K.}~\bibnamefont{Kubo}},
  \bibinfo{journal}{Phys. Rev. B} \textbf{\bibinfo{volume}{65}},
  \bibinfo{pages}{224403} (\bibinfo{year}{2002}).
  
  
\bibitem[{\citenamefont{Zaki}(2011)}]{Zaki}
\bibinfo{author}{\bibfnamefont{N.}~\bibnamefont{Zaki}} \bibinfo{note}{(private communication)}.

\bibitem[{\citenamefont{Carpentier and Orignac}(2006)}]{Carpentier2006}
\bibinfo{author}{\bibfnamefont{D.}~\bibnamefont{Carpentier}} \bibnamefont{and}
  \bibinfo{author}{\bibfnamefont{E.}~\bibnamefont{Orignac}},
  \bibinfo{journal}{Phys. Rev. B} \textbf{\bibinfo{volume}{74}},
  \bibinfo{pages}{085409} (\bibinfo{year}{2006}).

\end{thebibliography}
\end{document}